# Atmospheric Characterization via Broadband Color Filters
# on the PLAnetary Transits and Oscillations of stars (PLATO) Mission

John Lee Grenfell[1*], Mareike Godolt[2], Juan Cabrera[1], Ludmila Carone[3],
Antonio García Muñoz[2], Daniel Kitzmann[4], Alexis M. S. Smith[1] and Heike Rauer[1,2,5]

(1) Department of Extrasolar Planets and Atmospheres (EPA)
Institute for Planetary Research (PF)
German Aerospace Centre (DLR)
Rutherfordstr. 2
12489 Berlin
Germany

(2) Centre for Astronomy und Astrophysics (ZAA)
Berlin Institute of Technology (TUB)
Hardenbergstr. 36
10623 Berlin
Germany

(3) Max-Planck-Institute for Astronomy (MPIA)
Königstuhl 17
69117 Heidelberg
Germany

(4) Center for Space and Habitability (CSH)
Gesellschaftsstrasse 6
University of Bern
3012 Bern
Switzerland

(5) Institute for Geological Sciences
Planetology and Remote Sensing
Free University of Berlin (FUB)
Malteserstr. 74-100
12249 Berlin

*Contact author, email address: lee.grenfell@dlr.de

*Abstract: We assess broadband color filters for the two fast cameras on the PLAnetary Transits and Oscillations (PLATO) of stars space mission with respect to exoplanetary atmospheric characterization. We focus on Ultra Hot Jupiters and Hot Jupiters placed 25pc and 100pc away from the Earth and low mass low density planets placed 10pc and 25pc away. Our analysis takes as input literature values for the difference in transit depth between the broadband lower (500-675nm) wavelength interval (hereafter referred to as "blue") and the upper (675-1125nm) broadband wavelength interval (hereafter referred to as "red") for transmission, occultation and phase curve analyses. Planets orbiting main sequence central stars with stellar classes F, G, K and M are investigated. We calculate the signal-to-noise ratio with respect to photon and instrument noise for detecting the difference in transit depth between the two spectral intervals. Results suggest that bulk atmospheric composition and planetary geometric albedos could be detected for (Ultra) Hot Jupiters up to ~100pc (~25pc) with strong (moderate) Rayleigh extinction. Phase curve information could be extracted for Ultra Hot Jupiters orbiting K and G dwarf stars up to 25pc away. For low mass low density planets, basic atmospheric types (primary and water-dominated) and the presence of sub-micron hazes in the upper atmosphere could be distinguished for up to a handful of cases up to ~10pc.*

**Keywords**: PLATO, Color Filters, Atmospheres, Composition, Haze, Albedo

**1. Introduction**

The PLATO (PLAnetary Transits and Oscillations of Stars) space mission (Rauer et al. 2014) is the third medium-class (M3) mission in the European Space Agency (ESA) Science Cosmic Vision programme and is due for launch in 2026. PLATO will provide mean (bulk) planetary density and age of planetary-stellar systems including rocky planets in the Habitable Zone (HZ) around Sun-like stars at an unprecedented level of accuracy. The broadband color filters (blue: 500-675nm and red: 675-1125nm) on the fast cameras (FCs) will obtain photometric color information which could be linked to atmospheric properties. In this paper we investigate to what extent the FC color information can be used to constrain atmospheric parameters such as mean composition, haze properties and geometric albedo for Hot Jupiters (HJs) such as HD 209458b and lower mass planets such as GJ 1214b. We first summarize some key issues in the literature for HJs, Ultra HJs, and lower mass planets such as GJ 1214b. Our aim thereby is to strengthen the motivation for observing bulk composition, climate and albedo for these objects by the PLATO fast camera filters. To reach this goal, the PLATO filters will measure (1) the "Rayleigh absorption slope" in primary transit which constrains bulk composition and clouds, (2) the secondary transit depth varying with wavelength which constrains geometric albedo and outgoing emission, hence climate and (3) the phase curve which constrains clouds and transport.

HJ science has matured considerably in recent years. Studying these objects has extended our knowledge of planetary formation and migration (Mordasini et al., 2012; 2015; 2016; Batygin et al., 2016; see also Boley et al., 2016) as well as internal structure (e.g. Fortney and Nettelman, 2010) and chemical composition (Maldonado et al., 2018). Transmission and occultation spectroscopy has established the presence of numerous atmospheric species (see reviews by e.g. Tinneti et al. 2013; Madhusudhan, 2019) including water (e.g. Kreidberg et al., 2014[a]), titanium oxide (Sedaghati et al., 2017) as well as a discussion on potassium (Sing et al., 2015; Sedaghati et al., 2016; Gibson et al., 2017) and helium (Spake et al., 2018). Numerous works have discussed HJ climate and dynamics from occultation spectroscopy (e.g. Lee et al., 2012; Bean et al., 2013). The data suggests a broad continuum in HJ properties (Sing et al., 2016; Tsiaras et al., 2018) ranging from clear to hazy atmospheres. Whereas the clear atmospheres often exhibit alkali metal absorption, the hazy atmospheres have virtually featureless spectra at shorter wavelengths due to the masking effect of high altitude haze. No clear relation between e.g. planetary equilibrium temperature and the presence of such haze has however been found up to now (see e.g.

Sing et al., 2016; Huitson et al., 2017). Atmospheric retrieval of composition and temperature of HJs was discussed in e.g. Barstow et al. (2015) for the James Webb Space Telescope (JWST) and Pinhas et al. (2018). Recent progress on the HJ mass-radius anomaly - the so-called inflated HJs (see overview in e.g. Batygin and Stevenson, 2010; Laughlin, 2018) - suggested that inflation efficiency is mass and temperature dependent (Thorngren and Fortney, 2018; Bento et al., 2018).

Understanding of atmospheric transport on HJs (e.g. Amundsen et al., 2014, 2016; Heng and Showman 2015; Komacek and Showman, 2016) is also rapidly progressing. Drummond et al. (2018) suggested that coupling of 3D atmospheric dynamics and chemistry could be crucial in understanding meridional transport of key species such as methane. Regarding clouds, Sing et al. (2016) analyzed ten HJs and suggested a continuum of objects ranging from clear sky HJs with strong water absorption features, to cloudy HJs with weaker water absorption (see also Barstow et al., 2016[a]). Cloud composition variations with HJ equilibrium temperature were catalogued by Parmentier et al. (2016). Pinhas and Madhusudhan (2017) discussed cloud signatures in HJ transmission spectroscopy. Lines et al. (2018) discussed the ability of the JWST to distinguish HJ cloud properties.

Planetary phase curves of HJs contain information on stellar reflection (hence clouds and hazes), heat re-distribution efficiencies, thermal emission, Doppler relativistic beaming and stellar effects such as gravity-induced shape-modulations (see e.g. Močnik et al., 2018 and references therein). Since the earlier seminal studies (e.g. Cowan and Charbonneau, 2007) understanding has progressed rapidly although challenges in breaking the degeneracies (e.g. between clouds, transport and albedo, see below) remain. The phase curve data study by Zellem et al. (2014) supported earlier General Circulation Model predictions (Showman and Guillot, 2002) of equatorial superrotation producing an eastward shift in the sub-stellar point. A recent study of nine HJs (Kataria et al., 2016) suggested stronger superrotation with increasing equilibrium temperature.

Knowing the Rayleigh slope together with the molecular hydrogen abundance (estimated from the planetary mass and assuming this gas is the dominant species) constrains the pressure region over which the Rayleigh slope arises. This is useful information for interpreting the remaining spectral features (see section 3.0; Lecavelier des Etangs et al., 2008; Sing et al., 2008; Griffith, 2014). The Rayleigh slope also provides information on gas and haze composition although an important caveat is to allow for the effect of stellar plages (Osagh et al., 2014) which can mimic such a Rayleigh extinction feature. In the UV the Rayleigh feature is often affected by high-altitude, fine hazes and by the abundance of gas-phase species such as $H_2$, He and $H_2O$. Note that there are no significant absorption bands in this spectral region. Small haze particles can also produce "Rayleigh" scattering. The abundance of bulk species such $H_2$, He and $H_2O$ will affect the Rayleigh slope via the mean molecular weight. In the visible the absorption features of alkali metals (Sing et al., 2008) can be prominent for clear HJ atmospheres with low amounts of hazes and clouds. From the red end of the visible out to a few microns, strong absorption features can arise from water or/and metal oxides such as TiO and VO which can depend on e.g. the metallicity.

Determining the albedo of HJs provides fundamental insight into composition, clouds, energy budgets and global circulation (Sudarsky et al., 2000). Observed geometric albedos of HJs cover a wide range. Low albedos have been observed in the visible wavelength range (e.g. HD 209458b, $A_{geometric}$ = 0.038±0.045 for 400-700nm (Rowe et al., 2008); $A_{geometric}$ = 0.064 at $2\sigma$ (Bell et al., 2017) for WASP-12b; $A_{geometric}$ = 0.030 at $2\sigma$ (Močnik et al., 2018) for WASP-104b. The latter study suggested little or no dependence of albedo with wavelength, in contrast to studies of HD198733b (e.g. Berdyugina et al., 2011; Evans et al. 2013, appendix A3). Dark HJ albedos are consistent with a lack of clouds and hazes or/and efficient absorption in the visible by alkali metals. Bright HJs on the other hand have higher geometric albedos e.g. $A_{geometric}$ = 0.29-0.35 for Kepler-7b (Demory et al., 2011) and even higher for very hot Jupiters possibly due to silicate clouds (see e.g. Cowan and Agol, 2011[a]; Santerne et al., 2011 and references therein). Degeneracies exist between Bond albedo, equilibrium temperature and energy re-distribution efficiency (Cowan and Agol 2011[a]; Demory et al., 2011; Nikolov et al., 2018[a]) as well as in cloud properties and their location (e.g. Heng and Demory, 2013; García Muñoz and Isaak, 2015; von Paris et al., 2016). Schwartz and Cowan (2015) discuss additional data which could be gathered to break such degeneracies e.g. between observed albedo and the heat re-distribution efficiency. García Muñoz & Cabrera (2018) investigate the diagnostics possibilities of observing low-density exoplanets at large phase angles, since a number of them may show strong forward scattering by the prevailing upper atmosphere haze. García Muñoz (2018) furthers a technique to measure the polarization signal of starlight reflected by exoplanets that complements brightness-only measurements. He discusses constraints that could be set for the cloud and haze particles (e.g. their size, shape and refractive index). The PLATO filters provide the opportunity of having spectrally-resolved data from transit to occultation in the visible wavelength range. This, although more challenging to obtain than in the IR due to the stronger accuracy requirements is nevertheless feasible (see e.g. Désert et al., 2011) and can then be combined with IR data to address such degeneracies and hence to better constrain climate.

HD 209458b (Charbonneau et al., 2000) with $R_{HD\ 209458b}=1.35R_J$ and $M_{HD\ 209458b}=0.71M_J$) is one of the most well-studied HJs. It transits a G0V star with an orbital period of 3.52 days and is located 48.9 pc away in the Pegasus constellation. It was the first exoplanetary target with empirical evidence of an atmosphere via spectroscopic detection of sodium (Charbonneau et al., 2002); an evaporating hydrogen tail (Vidal-Madjar et al., 2003); atomic oxygen and carbon (Vidal-Madjar et al., 2004); an ongoing discussion on water detection (e.g. Swain et al., 2009; Line et al., 2016); orbital speed and planetary winds (Snellen et al., 2010) as well as indications of a magnetic field (Kislyakova et al., 2014). Recent studies with updated observational reduction methods (Diamond-Lowe et al., 2014) or/and at higher wavelength resolutions (e.g. Line et al., 2016) suggested no temperature inversion on HD 209458b, in contrast to earlier photometric studies (e.g. Knutson et al., 2008). This has therefore fueled debate as to whether aerosol species such as VO and TiO - which are associated with temperature inversions - are too heavy to be maintained at sufficiently high altitudes (see e.g. discussion in Diamond –Lowe et al., 2014) on HD 209458b. Clouds on HD 209458b were investigated by applying e.g. sophisticated models (e.g. Helling et al., 2016) or/and Bayesian analyses (McDonald and Madhusudhan, 2017). Numerous 3D model studies have investigated transport on HD 209458b (e.g. Showman et al., 2009; Drummond et al., 2018). Studies investigating the Rayleigh slope seen in transmission spectra for HD 209458b and other HJs are discussed in section 6 which assesses the signal of the Rayleigh slope, the secondary transit depth and the phase curve amplitude compared with expected photon and instrument noise of the PLATO filters measurements. Due to the well-studied nature of HD 209458b we adopt its planetary and atmospheric parameters for our S/N calculations (see section 6.0).

Ultra Hot Jupiters (UHJs) (defined as HJs with orbital periods of ~1day or shorter and $T_{eq}>2000K$) represent a newly-emerging class of planets (see e.g. Hebb et al., 2009[a,b] who reported the discovery of WASP-12b and WASP-19). Earlier studies e.g. Sing and López-Morales (2009) and Gillon et al. (2012) observed occultation eclipses on UHJs. UHJs feature e.g. cycling between molecular thermolysis in hotter regions and exothermic recombination of atmospheric species in cooler regions (see e.g. Bell and Cowan, 2018; Kitzmann et al., 2018). Some studies e.g. Stevenson et al. (2014) suggest that the UHJ WASP-12b could be carbon-rich with C/O>1 although this is still discussed (Kreidberg et al., 2015). Smith et al. (2011) suggested a brightness temperature of 3620K for WASP-33b at 0.91 microns. Occultation observations for the UHJ WASP-103b (Cartier et al., 2018) in the NIR suggested a brightness temperature of 2890K. Their study noted that having additional data in the visible (e.g. delivered by the PLATO FCs) would help constrain metallicity, clouds and temperature profiles. Eclipse photometry performed on WASP-103b (Delrez et al., 2018) suggested a dayside temperature of ~2900K and an optical transmission spectrum with an observed minimum at ~700nm. Thermal phase curve data from HST and Spitzer for this planet (Kreidberg et al., 2018) suggested a (dayside only) thermal inversion with a lack of spectral water features possibly due to thermal dissociation whereas phase curve data could place constraints on the planetary magnetic field. Hoeijmakers et al. (2018) suggested iron and titanium in the atmosphere of the UHJ KELT-9b from high-resolution spectroscopy. Evans et al. (2018) suggested that the UHJ WASP-121b featured a stratosphere, possibly due to high-altitude clouds. Arcangeli et al. (2018) suggested a transition region at ~2500K above which $H^-$ opacity becomes a significant feature in the spectra. In our work we adopt the observed planetary parameters of the UHJ WASP-103b namely $R_{WASP-103b}$=1.528$R_J$; $M_{WASP-103b}$=1.49$M_J$ and $T_{eq}$=2508K (Gillon et al., 2014) for our S/N calculations of transit occultation (see section 6) for a hypothetical planet placed at 25pc and 100pc.

On moving to lower planetary masses, some warm and hot Neptunes have been characterized with photometry and spectroscopy e.g. for GJ 3470b with $R_{GJ\ 3470b}$=3.88$R_E$ and $M_{GJ\ 3470b}$=12.58$M_E$ (Awiphan et al., 2016; Kosiarek et al., 2019) (by comparison $M_{Neptune}$=17.15 $M_E$) and for HAT-P-11b with $R_{HAT-P-11b}$=4.31$R_E$ and $M_{HAT-P-11b}$=25.8$M_E$ (Bakos et al., 2010; Huber et al., (2011)[a,b]; Demory et al., 2011; see also Mansfield et al., 2018). At lower masses, Fortney et al. (2013) discussed the class of exoplanets including GJ 1214b (hereafter GJ 1214-like objects, GLOs) in the range 2-4 Earth radii and ~<30 Earth masses. Satellite data from the Kepler mission (Borucki et al., 2010) suggested that these objects should be rather common (Howard et al., 2010).

Key issues when studying GLOs are to determine to what extent they can (a) retain their primordial, light atmospheres, or (b) have thick water rich atmospheres [or form mixtures of (a) and (b)]. Venturini et al. (2016) presented an updated planetary formation model including the effect of envelope enrichment and noted good agreement with observed water abundances for GLOs (and HJs). Determining the GLO radius from theory depends upon knowledge of e.g. the equation of state (e.g. Lopez and Fortney, 2014) in the interior. Recent GLO formation studies (Chachan and Stevenson, 2018) assuming thick $H_2$-envelopes with rocky cores noted the importance of including $H_2(g)$ dissolution into the interior; this gas is then gradually outgassed which leads overall to a significant slowing in integrated atmospheric loss rates. Dorn et al. (2017) constrained planetary ice and rock mass fractions using a Bayesian analysis based on stellar abundance proxies. The Bayesian analysis of Dorn and Heng (2018) suggested mainly secondary (non-hydrogen) atmospheres on the cool SEs HD 219134b.

GJ 1214b (Charbonneau et al., 2009) ($R_{GJ\ 1214b}$=2.7$R_{Earth}$ and $M_{GJ\ 1214b}$=6.6$M_{Earth}$) transits a star bright enough for both the planetary radius and mass to be determined and is therefore relatively well-studied. Transmission spectroscopy rules out a cloud-free, hydrogen atmosphere (Kempton et al., 2012; Fraine et al., 2013; de Mooij et al., 2013; Kreidberg et al., 2014[b]). Regarding model studies of GJ 1214b, Fortney et al. (2013) applied a population synthesis model which suggested a high metallicity of hundreds of times the solar value. Valencia et al., (2013) applied an evolutionary model to GJ 1214b which suggested an upper limit of 7% for the mass of an H/He envelope. Kempton et al., (2012) and Hu and Seager (2014) discussed photochemical effects for GJ 1214b using an atmospheric column model. The atmospheric circulation of GJ 1214b has been studied using a simplified 3D model with grey radiative transfer scheme (Menou, 2012) and with full General Circulation Models (GCMs) (Kataria et al., 2014; Charnay et al., 2015[a,b]). Gao and Benneke (2018) recently applied a cloud model to GJ 1214b postulating the presence of potassium chloride and zinc sulphide clouds. Lavvas et al. (2019) studied the properties of potential hazes on GJ 1214b. Factors affecting the Rayleigh slope for GJ 1214b (and other GLOs) are discussed in section 6.3. Due to the well-studied nature of GJ 1214b we adopt its planetary and atmospheric parameters for our S/N calculations. At present there are many open questions regarding the nature of GJ 1214b and detailed spectral measurements are rather lacking. Nevertheless, in our work we will use the numerous modeling studies of this planet as a basis for our analysis.

Super-Earths (SEs) are rocky worlds with masses less than ~10 Earth masses and radii commonly less than ~2 Earth radii. The mass definition is based on theoretical estimates of the critical mass for a nucleated instability e.g. Rafikov (2006) leading to rapid $H_2$/He accretion – it is however only approximate and depends strongly e.g. on planet-star distance. Photoevaporation of hydrogen and helium could transform GLOs into SEs (Luger et al., 2015; see also Jin and Mordasini, 2018). Horst et al. (2018) characterized simulated atmospheres over a broad (T, p, composition) range for SEs and GLOs and suggested great diversity in expected haze conditions. Grenfell et al. (2018) proposed that explosion-combustion limits the range of $H_2$-$O_2$ compositions on SEs. Southworth et al. (2017) discussed possible atmospheric spectral features detected on GJ 1132b. The SE 55 Cancri e (Fischer et al., 2008) ($R_{55Cancri\,e}$=2.0$R_{Earth}$; $M_{55Cancri\,e}$=8.4$M_{Earth}$ Endl et al., 2012) has been suggested to consist of carbon-rich material possibly with a water-rich envelope (but see also Crida et al., 2018); CoRoT-7b ($M_{CoRoT-7b}$ =7.4$M_{Earth}$; Hatzes et al., 2011; $R_{CoRoT-7b}$ =1.585$R_{Earth}$; Barros et al., 2014), Kepler 10b (Batalha et al., 2011) ($R_{Kepler10b}$=1.4$R_{Earth}$; $M_{Kepler10b}$=4.6$M_{Earth}$) and K2-265b (Lam et al., 2018) ($R_{K2-265b}$=1.71$R_{Earth}$; $M_{K2-265bb}$=6.54$M_{Earth}$ are examples of rocky, hot SEs. The number of cool SE and sub-Earth rocky exoplanets known is also rapidly expanding. For example, the nearby ultra-cool (M8V) M-dwarf TRAPPIST-1 (TRAnsiting Planets and Planetesimals Small Telescope) spectral system was found to feature seven temperate terrestrial planets with ~Earth-like masses (Gillon et al., 2017) which transit their star. First hints on atmospheric composition are emerging e.g. a cloud-free, hydrogen dominated atmosphere is ruled-out for TRAPPIST d,e and f (de Wit et al., 2016; 2018). Gandolfi et al., (2018) and Huang et al. (2018) recently reported the discovery of the first exoplanet found by the Transiting Exoplanet Survey Satellite (TESS) mission, namely the SE, $\pi$ Men c.

## 2.0 Motivation

We will determine the extent by which the optical broadband filters in the fast cameras of the PLATO mission can constrain atmospheric composition, cloud/haze properties and planetary albedo of (U)HJ and GLO planets orbiting different main sequence stars placed up to 100pc away from the Earth.

## 3.0 Theory

### 3.1 Deriving Atmospheric Parameters from the Rayleigh Slope

The transit depth difference ($\Delta$TD) for a planet with an atmospheric contribution minus the same planet but without the atmospheric contribution can be written (see e.g. Kreidberg, 2018):

$$\Delta TD = [(R_p + nH)^2 / R_*^2] - (R_p^2 / R_*^2) \qquad (1)$$

$R_p$=planetary radius without atmospheric contribution; H=atmospheric scale height, n=number of scale heights typically sampled where n has typical values ranging from about 1 to 5 (for cloud-free atmospheres at low resolution) and $R_*$=stellar radius. Griffith (2014) (e.g. their equation 3.4) discuss the pressure region and number of scale heights sampled by primary transit. Assuming that $\Delta TD_{blue}$ and $\Delta TD_{red}$ denote TDs in the FC blue and red intervals respectively, expanding (1) and neglecting quadratic terms:

$$\Delta TD_{blue} - \Delta TD_{red} \approx 2nH\,R_p / R_*^2 \qquad (2)$$

Appendix A1 shows (a) calculated 'n' (number of atmospheric scale heights) from (2) as sampled by the FC blue and red intervals and (b) the corresponding vertical range sampled, in order to interpret the observations and compare with model studies for different atmospheric compositions. For HD 209458b, appendix A1 suggests that the FC cameras would sample n~0.3 scale heights at ~97% of the planet's radius. For GJ 1214b they would sample n~1.1 scale heights at ~92% of the planet's radius. Here, the lower scale height of the water-dominated atmosphere compared with the x0.01 Solar atmosphere approximately cancels the effect of the weaker $TD_{blue-red}$ of the water case, so that overall the two cases yield a similar value for number of scale heights sampled (n~1.1 in both cases). For an Earth-like planet orbiting AD Leo, appendix A1 suggests that the FCs would sample across ~0.7 scale heights in the troposphere and for the Earth case they would observe over ~1.1 scale heights in the lower stratosphere. The atmospheric scale height is defined:

$$H = k_B T/\mu g \qquad (3)$$

$k_B$=Boltzmann's constant, T=temperature, $\mu$=mean atmospheric molecular mass and g=gravitational acceleration. Earlier studies (e.g. Miller-Ricci et al., 2008) discussed deriving H in equation (3) via IR transmission spectroscopy. In the IR the H value is degenerate depending on cloud and haze properties and distribution (Benneke and Seager, 2012) and the (possible) presence of a surface. The temperature is typically assumed to be either the planetary equilibrium temperature, $T_{eq}$ or is calculated from a radiative transfer model.

To address these difficulties, a revised determination of H (Benneke and Seager, 2012) was proposed in the optical/ultra-violet (UV) region assuming that Rayleigh scattering dominates extinction along the line of sight. Firstly, assuming (Lecavelier des Etangs et al., 2008):

$$H = dR_{p,\lambda}/d(\ln\sigma_\lambda) \qquad (4)$$

$\sigma_\lambda$=Rayleigh scattering cross-section. Then, combining (3) and (4), substituting: $\sigma(\lambda) \propto \lambda^{-4}$ (Rayleigh extinction) (at wavelengths $\lambda_1$ and $\lambda_2$ at temperatures $T_1$, $T_2$ respectively) and re-arranging yields:

$$\mu = (4k_bT/gR_*) \ [\ln(\lambda_1/\lambda_2)/[(R_p/R_*)_{\lambda 2} - (R_p/R_*)_{\lambda 1}] \ * \ (1\pm\delta T/T) \qquad (5)$$

Equation (5) highlights here $\delta T$, the uncertainty in T - although clearly uncertainty in all its variables, namely g, $R_p$, $R_*$ should also be considered. Equation (5) states that by measuring the transit depth at two wavelengths which are dominated by Rayleigh scattering one can constrain the atmospheric mean molecular mass ($\mu$). The Rayleigh extinction feature ("Rayleigh slope") in equation (5) is not subject to all the degeneracies discussed above although these can arise if the atmosphere contains multiple main components (e.g. $N_2$-$O_2$); also, hazes can influence the Rayleigh slope as discussed in Benneke and Seager (2012), see also Griffith (2014) and Heng, (2016); note that a given haze may not display a Rayleigh scattering behavior at all depending on haze properties.

The change in $\sigma_{rayleigh}(\lambda)$ (the Rayleigh scattering cross-section for atoms or molecules) from equation (4) over the wavelengths of the PLATO fast cameras for the benchmark case of a pure $H_2(g)$ cloud-free atmosphere can be calculated from Rayleigh theory which states that $\sigma_\lambda$ depends on wavelength and refraction index (n) as follows:

$$\sigma_{rayleigh}(\lambda) = (24\pi^3\nu^4/n^2_{ref}) \ [(n(\nu)^2-1)/ \ (n(\nu)^2+2)]^2 \ K(\nu) \qquad (6)$$

where $\nu$ is wavenumber, n is the refractive index, $n_{ref}$ is a reference particle number density and K is the King factor, a measure of the non-sphericity of the molecules. Note that Rayleigh scattering can be dominated in some cases by a small number of fine haze particles (Pont et al., 2013). For a cloud-free, pure $H_2(g)$ atmosphere, Rayleigh scattering (Vardya, 1962 and references therein) can be approximated by:

$$\sigma_{rayleigh}(\lambda) = C/\lambda^4 = (8.4909 \times 10^{-45} cm^6/\lambda^4) \qquad (7)$$

where $\lambda$ has the units cm. Substituting the PLATO fast camera blue (=5.88x10$^{-5}$ cm) and red (=9.00x10$^{-5}$ cm) wavelength midpoints into equation (7) leads to a change by more than a factor of five in the Rayleigh cross-section, $\sigma(\lambda)$, since $(\sigma_{blue}/\sigma_{red}) = [\lambda_{red}/\lambda_{blue}]^4 = 5.51$. Such (atmospheric) effects have however only a small effect on the observed transit depth since the (non-optically thick) portion of the atmosphere which is sampled by transmission spectroscopy typically represents only a small fraction of the total planetary radius. The impact of the Rayleigh extinction feature therefore results in typically only up to ~one tenth of a percent increase in the observed primary transit depth in the "blue" compared to the "red" wavelengths (see Table 3).

### 3.2 Deriving Geometric Albedo from Reflection Spectroscopy

By measuring the flux difference at different wavelengths in and out of occultation one can calculate (Rowe et al. 2008; their equation 2) the geometric albedo of the atmosphere via:

$$F_p(\lambda) = (R_p^2/a^2) A_g(\lambda) F_* \qquad (8)$$

$F_p$=reflected planetary flux, a=planet-star orbital distance, $A_g$=geometric albedo, $F_*$ = photon flux at the stellar surface. For HD 209458b, substituting $R_p$=1.35$R_J$ , a=7.10x10$^6$km yields:

$$F(\lambda)_{HD\,209458b} = 1.85 \times 10^{-4} A_g(\lambda) F_* \qquad (9)$$

(assuming the signal is dominated by reflected starlight; hotter planets will of course have stronger contributions from thermal emission, as discussed below).

### 3.3 Deriving Atmospheric Properties of UHJs from Optical Phase Curves

The new class of Ultra Hot Jupiters ($T_{eq}$>2000K) are considerably brighter in the optical (as discussed in 6.2) than HJs. Although optical phase curves are more challenging to obtain than those in the IR, they deliver important additional information which addresses e.g. cloud, albedo and transport degeneracies. Section 7.0 calculates the S/N for primary transmission, occultation and the direct emission of optical flux in the red and blue PLATO FC intervals for an UHJ with the assumed planetary properties of WASP-103b placed 25pc and 100pc from the Earth orbiting F, G and K stars.

## 4.0 Method

We take wavelength-dependent transit depths from theoretical atmospheric transmission spectra published in the literature binned into the PLATO broadband filter wavelength intervals, hereafter referred to as the "blue"$_{500\text{-}675nm}$ and "red"$_{675\text{-}1125nm}$ intervals. We then compare the statistical significance of the broadband-binned transit depths for representative levels of photon noise plus instrument noise (see 5.3) assumed to be equal to 30% of the photon noise.

## 5.0 Mission Concept, Camera Design and Noise Analysis

## 5.1 Mission Concept

The overall scientific goals of the PLATO mission aim at answering the following questions: how do planets and planetary systems form and evolve? Is our solar system special or there are other systems like ours? Are there potentially habitable planets? PLATO will survey up to about half of the sky (Rauer et al., 2014) with a total field of about 2250 degrees$^2$ per pointing. The current baseline case for the four year nominal science mission will observe two sky fields for two years each over two long pointings. An alternative observing framework currently under discussion is to perform a long pointing lasting three years followed by a one year step-and-stare phase with several pointings. The exact observing strategy will be decided 2 years before launch and can be adapted flexibly. Mission extensions are possible up to the end of consumables after 8 years.

## 5.2 Camera Design

PLATO features 24 'normal' cameras with cadences of 25s targeting the fainter stars with visible magnitudes, $M_v>8$ plus two 'fast' cameras with cadences of 2.5s targeting the brighter stars with ~$4<M_v<8$. Each of the cameras, both normal and fast, has a pupil size diameter of 12cm. A detailed technical description can be found in Ragazzoni et al. (2016). Each PLATO camera features four, specifically-designed large-format (8x8cm) thin gate, back-illuminated CCDs having (4510x4510) 18μm square pixels. The CCDs push current technological frontiers in terms of sensitivity, precision, readout speed and full well capacity (see Prod'homme et al., 2016 for more details). Marcos-Arenal et al. (2014) simulated CCD performance, saturation and noise using the PLATO SIMULATOR software package at the nominal working temperature of -80°C. The PLATO fast cameras are responsible for delivering the fine guidance information to the attitude and orbital control system of the spacecraft. There are two fast units to fulfill the reliability goals of the mission and each observes in a different wavelength interval. For redundancy, the wavelength intervals were chosen in such a way that each camera collects 50% of the flux of a G0 star, which were taken as the reference stars for the mission performance. The PLATO fast cameras aim for a spectral overlap of ≤30% with a total throughput of ≥85% in each band, fulfilling the requirements for fine guidance and high precision photometry (see also Verhoeve et al., 2016).

### 5.3 Noise Analysis

The PLATO mission is designed to achieve total noise levels of less than 50 ppm in 1h for $m_V$<11 magnitude stars. Instruments design is such that the sum of non-photonic noise sources remains below ~one third of the photon noise. Pointing noises such as jitter are minimized by placing the instrument in the stable L2 orbit and by using the two fast cameras to deliver high frequency pointing error information to the satellite's attitude and orbit control system. Jitter noise seen as displacements of star images on the detector also undergo on-ground correction through knowledge of the point spread function.  Long term drift in the CCD bias voltage is reduced by a first stage filtering inside the ancillary electronic units and a second stage filtering in the front end electronics. Background noise from straylight and dark current is minimized by the telescope optical design with its nominal temperature of -80°C at a reference point close to the pupil plane and by a baffle placed in front of each camera. Each camera has its own detector readout electronics. A detailed discussion of the PLATO 2.0 noise assessment can be found in Laubier et al., (2017).

To estimate the photon noise, we calculate fluxes entering the fast cameras by integrating the Planck black body radiation from (500-1120nm) in 50nm intervals for a given stellar temperature, radius and distance from the Earth. We assume in our analysis a transit time of one hour as a first estimate and in order to facilitate comparison between the stellar scenarios. The effect upon our calculation of, for example doubling the assumed transit time from one to two hours (assuming a Poisson distribution) would imply a reduction in the noise by a factor √2. A caveat is that stars do not radiate exactly as blackbodies although this effect is small for our purposes. For calibration we assume that the flux received from an $M_v$=0 star is $3.6182 \times 10^{-12}$ W cm$^{-2}$ micron$^{-1}$ in the Johnson-Cousins V-band at 550nm (Casagrande and VandenBerg, 2014). The collecting surface corresponds to an entrance pupil diameter of 12cm assuming for the fast cameras 2.3s exposure plus 0.2s frame transfer for a cadence time of 2.5s. The instrument point spread function (PSF) is designed such that 90% of the incoming energy for a given target star falls within a 2x2 pixel square. Final photon fluxes are integrated over one hour assuming that the PSF is positioned so that the energy deposition is maximum. We assume a hypothetical planet with a transit time of one hour in all cases. For comparison, $t_{transit}$~53 minutes for GJ 1214b (Berta et al., 2012); ~90 minutes for WASP-19b (a UHJ) (Sedaghati et al., 2015) and ~184 minutes for HD 209458b (a HJ) (Tsiaras et al., 2016). There are more detailed noise simulations for the PLATO instrument available (see Samadi et al., 2019). However, we use this simplified approach since firstly, it is accurate enough for our purposes and secondly, since our noise estimates are consistent with the more detailed simulations.

### 5.3.1 Photon and Instrument Noise

The photon noise (PN) is calculated from the final, time-integrated incoming photon flux ($F_{phot}$) assuming that the incoming photons follow a Poisson distribution. Expressing as a fraction of the final incoming photon flux and converting to parts per million (ppm), the 1-sigma standard deviation due to PN can be expressed as:

$$\sigma_{PN} = 10^6 * (\sqrt{F_{phot}}/F_{phot}) = 10^6/\sqrt{F_{phot}} \text{ (ppm)} \tag{10}$$

In addition to the photon noise, there also features instrument noise (due to readout etc.). In our calculations we assume that the joint error due to photon plus instrument noise is equal to 1.3 times the photon noise (see equation 10) according to the PLATO mission noise requirement (see e.g. Rauer et al., 2011).

### 5.3.2 Error Analysis

The aim is to calculate the statistical significance of the difference in Transit Depth (TD) for the blue and red spectral intervals i.e. (TD$_{blue}$-TD$_{red}$) where:

$$TD = \delta F = (F_{out} - F_{in}/F_{out}) \tag{11}$$

$F_{out}$ and $F_{in}$ denote stellar fluxes for out-of-transit and in-transit respectively. From the theory for the propagation of non-linear errors (see e.g. Bevington et al., 2002), the error ($=\sigma(\delta F)$) of the flux difference ($F_{out}-F_{in}$) relative to the background flux, $F_{out}$ is given by:

$$\sigma(\delta F) \approx \sqrt{\left(\frac{\delta(Fout)}{Fout}\right)^2 + \left(\frac{\delta(Fin)}{Fin}\right)^2} \tag{12}$$

On assuming $F_{out} \sim F_{in}$ and $\delta F_{out} \sim \delta F_{in}$, then equation (12) simplifies to:

$$\sigma(\delta F) \sim \sqrt{2}(\delta F/F) \tag{13}$$

Similarly, the error ($=\sigma(F_{blue}-F_{blue})$) associated with the flux difference between the two wavelength intervals ($F_{blue}-F_{red}$) can be written:

$$\sigma(F_{blue}-F_{red}) \approx \sqrt{\left(\frac{\sqrt{2}\delta(Fblue)}{Fblue}\right)^2 + \left(\frac{\sqrt{2}\delta(Fred)}{Fred}\right)^2} \tag{14}$$

On assuming $F_{blue} \sim F_{red}$ and $\delta F_{blue} \sim \delta F_{red}$, then equation (14) simplifies to:

$$\sigma(F_{blue}-F_{red}) \approx 2(\delta F/F)^2 \tag{15}$$

Note that photon noise, instrument noise, and instrument specifications such as CCD transmission and quantum efficiency are taken into account in the term "$\delta F$" in equation (15).

### 5.3.3 Limb darkening effects

Uncertainty associated with limb darkening arises due to challenges in determining e.g. (1) the limb darkening coefficients (LDCs) and (2) the stellar class or sub-class, upon which the LDCs depend. Regarding (1), Csizmadia et al. (2013) suggested differences of up to ~30% in both the first two ($u_a$ and $u_b$) LDCs depending on e.g. the stellar model employed (e.g. PHOENIX or ATLAS), stellar temperature and metallicity. Regarding (2), stellar temperatures are typically uncertain by up to a few hundred Kelvin depending upon e.g. stellar class, stellar magnitude and the measurement technique (see e.g. Berriman et al., 1992). For example a mid-range M-dwarf with an assignment of e.g. M4.5 in stellar class could be typically uncertain from ~(M3.5-M5-5). Csizmadia et al., (2013) (their Figure 1, upper panel) suggested the increasing temperature by 200K from e.g. (3800-4000K) in stellar temperature corresponds to a change in $u_a$ and $u_b$ by 39% and 26% respectively. In our analysis we take (see Table 1) a lower and upper limit for observed blue minus red TDs (i.e. 50ppm and 500ppm) which is a large range (x10) compared with the uncertainty due to limb darkening (several tens of percent). This suggests that our analysis is representative, including the effects of limb darkening.

Limb darkening effects for stellar spectral types ranging from M to F are discussed in Claret (2018). Table 1 shows the estimated percentage by which Limb Darkening (%LD) can influence the derived radius ratio (=$R_p/R_*$)$^2$ of transiting planets for the three (F,G,K) stellar scenarios taken from Csizmadia et al. (2013):

Table 1: Theoretical linear ($u_a$) and quadratic ($u_b$) LDCs averaged over the CoRoT white light passband interval derived from models shown in Figure 1 of Csizmadia et al. (2013). [#]Flux ratio including limb darkening, $\Delta F/F = (\frac{Rp}{R_*})^2/[1- u_a/3 - u_b/6]$ for central transits (Csizmadia et al., 2013; their Figure 1, middle panel, red lines). %LD = %Change in flux ratio [=($\Delta F/F$ - 1.0)*100%] on including limb darkening.

| Star | $u_a$ | $u_b$ | $\Delta F/F$[#] | %LD |
|---|---|---|---|---|
| K5V (4130K) | 0.61 | 0.13 | $1.29(R_p/R_*)^2$ | 29% |
| Sun (5800K) | 0.40 | 0.25 | $1.21(R_p/R_*)^2$ | 21% |
| F5V (6540K) | 0.30 | 0.28 | $1.17(R_p/R_*)^2$ | 17% |

Table 1 suggests that the planetary radius would be under-estimated by several tens of percent if limb darkening is neglected, with strongest %LD values for the cooler stars. Furthermore, LDCs can be wavelength-dependent. Hestroffer and Magnan (1998) suggested for example that the center to limb darkening flux coefficients for the Sun increased by ~x2 with increasing wavelength over the PLATO FC wavelength range. Limb darkening (and other relevant effects such as stellar spots) and their colour dependencies are treated to different extents in the literature. In the present work, we take the TD values directly from the literature (see below).

### 5.3.4 Stellar spots

The presence of stellar spots can mimic the signature of Rayleigh absorption (see e.g. Mallonn et al., 2018; Oshagh et al., 2014) and their effects have to be removed when preparing the transmission spectra. This can be achieved by estimating plage distribution and activity from stellar models or by including stellar spots as a component of the red noise. Note that Parviainen et al., (2016) (Table 3) observed a large Rayleigh absorption feature which their atmospheric models could not explain. Most works however cited in Table 3 included the effects of stellar activity and could account for the Rayleigh slopes observed. This suggests that the lower and upper limits for TD taken from Table 3 for our S/N analysis is a representative range.

### 6.0 Input

### 6.1 Hot Jupiters

### 6.1.1 Planetary and Stellar Properties

We consider nominal HJs with the same radius and instellation as HD 209458b assuming a solar composition for its atmosphere and placed at distances of 25pc and 100pc from the Earth. For the G5, K5 and F5 scenarios we position the planet at an orbital distance where it receives the same prescribed stellar instellation as HD 209458b then calculate the orbital period from Kepler's 3$^{rd}$ Law (see Table 2). The assumed stellar and planetary properties are shown in Table 2 for the G5V, K5V and F5V cases. Table 2 also shows stellar class G0V (corresponding to HD209458) for comparison. The TD (see Table 3) is adjusted (in both the blue and red intervals) taking into account the different stellar radii relative to the G0V case. We do not consider HJs orbiting M-stars since these objects are rare (although there are exceptions, e.g. KOI-254, see Johnson et al., 2012).

Table 2: Stellar and planetary properties for Hot Jupiter scenarios. Stellar properties shown are class, luminosity ($L_*$), temperature ($T_*$), Mass ($M*$), radius ($R_*$) and visible magnitude ($M_v$). Planetary properties shown are radius ($R_p$), distance from star ($d_p$) in astronomical units (au) and orbital period ($P_p$) and are based on HD 209458b (which orbits a G0V-star) and its analogues orbiting other stars with planets placed so they receive the same stellar instellation.

| Class | $L_*$ (L$_{sun}$) | $T_*$ (K) | $M_*$ (M$_{sun}$) | $R_*$ (R$_{sun}$) | $M_v^*$ | $R_p$ (R$_{jupiter}$) | $d_p$ (AU) | $P_p$ (d) | Reference(s) |
|---|---|---|---|---|---|---|---|---|---|
| G0V | 1.61 | 6000 | 1.13 | 1.14 | 6.41 (25pc) 9.42 (100pc) | 1.35 | 0.045 | 3.52 | Boyajian et al., (2014) Mazeh et al. (2000) |

| G5V | 0.78 | 5520 | 0.92 | 0.93 | 7.07 (25pc) 10.08 (100pc) | 1.35 | 0.031 | 2.09 | See text Allen and Cox (2000) |
| K5V | 0.16 | 4130 | 0.69 | 0.74 | 9.25 (25pc) 12.26 (100pc) | 1.35 | 0.014 | 0.72 | Allen and Cox (2000) |
| F5V | 2.5 | 6540 | 1.3 | 1.207 | 5.58 (25pc) 8.59 (100pc) | 1.35 | 0.056 | 4.31 | 'Eta Arietis' Cenarro et al. (2007) |

Table 2 suggests that the orbital period of a planet with the radius of HD 209458b would increase by a factor of ~5.7 on moving from the cooler (K5V) to the hotter (F5V) case.

**6.1.2 Primary Transit**

For our calculations we seek representative, primary transit depth value(s) averaged over the PLATO fast camera blue and red wavelength intervals. Table 3 summarizes the range of observed HJ transit depths taken from the literature based on observations of HJ transmission spectra. The significance of the $TD_{blue-red}$ values in Table 3 depends on the uncertainties of the individual datapoints. In Table 3 (see legend for further details), the values of $TD_{blue-red}$ shown arise either from (1) data, where we calculated a linear average of all datapoints equidistant in wavelength lying in the blue or red intervals, or (2) from atmospheric model studies in the literature, where data was sparse in the blue and red wavelengths or problematic (e.g. due to an uncertain outlier). The atmospheric composition assumed in these models was constrained where possible using data in the IR and near IR (where data in the blue and red intervals was sparse). Values in Table 3 for model-based cases were calculated from the midpoints of the blue and red intervals lying on the line of spectral absorptions from the model output. The model output we chose in the studies shown achieved the best-fit to data taken over all wavelengths observed, extending typically from the UV/visible up to the mid-infrared, (and not just the blue and red PLATO filter intervals). Further details of the individual studies, the data and the models used, are shown in the text below Table 3.

Table 3: Primary transit depths (TDs) (ppm) averaged over the PLATO fast camera wavelength intervals blue (500-675nm, $TD_{blue}$), red (675-1125nm, $TD_{red}$) (unless stated otherwise) and their difference ($TD_{blue-red}$) for a range of HJs, the hot Saturn WASP-29b and the warm Saturn WASP-39b taken from the literature.

| $TD_{blue}$ (ppm) | $TD_{red}$ (ppm) | $TD_{blue-red}$ (ppm) | HJ (stellar class) | Reference(s) |
|---|---|---|---|---|
| 24,398 | 24,180 | 218 | HD 187933b (K1.5) | Sing et al. (2011)[1]* Figure 14 |
| 24,555 | 24,149 | 406 | HD 187933b | Pont et al. (2013)[+] Figure 11 dotted line |
| See text | See text | ~0 | HD 187933b | McCullough et al. (2014)[^] |
| 24,586 | 24,336 | 250 | HD 187933b | Angerhausen et al. (2015)[%] Figure 7 |
| 9,448 | 9,448 | ~0 | WASP-29b (K4) | Gibson et al. (2013)[a°] Figure 8 |
| 22,922 | 22,922 | ~0 | HAT-P-32b (F/G) | Gibson et al. (2013)[b°°] Figure 11 |
| 22,530 | 22,410 | 120 | HAT-P-32b | Mallonn and Strassmeier (2016)[°°°] Figure 10b, red line |
| 14,630 | 14,580 | 50 | HD 209458b (G0) | Deming et al. (2013)[**] Figure 14 |
| 14,003 (13,806) | 14,066 (14,066) | -63 (-260) | HAT-P-1b (G0) | Nikolov et al. (2014)[#] Figure 14 |
| 21,229 (21,170) | 20,982 (20,982) | 247 (188) | WASP-6b (G8) | Nikolov et al. (2015)[##] Figure 10 |
| 19,182 | 18,906 | 276 | WASP-6b | Jordán et al. (2013)[^^] Figure 10, green line |
| ~21,200 | ~21,200 | ~0 | WASP-39b (G8) | Nikolov et al. (2016)[###] Figure 4 Wakeford et al. (2018) |
| 27,298 | 27,298 | ~0 | WASP-31b (F6) | Sing et al (2015)[`] Figure 12, purple line |
| See text | See text | See text | WASP-17b (F6V) | Bento et al. (2014)[++] Sedaghati et al. (2016) Sing et al. (2016) |
| See text | See text | ~0 | HAT-P-12 (K5) | Mallonn et al. (2015)[%%] Figure 6, green line |
| 26,830 (26,896) | 25,664 (25,664) | 1,173 (1,232) | TrES-3b (GV) | Parviainen et al. (2016)[$] Figure 15 |
| 27,889 | 27,523 | 366 | TrES-3b | Mackebrandt et al. (2017)[%%%] Figure 9, black dashed |
| 19,432 | 18,769 | 663 | HAT-P-18b (K2) | Kirk et al. (2017)[$$] Figure 5c, green line |
| See text | See text | ~0 | WASP-4b (G8) | Huitson et al. (2017)[&] Figure 11, pink line |
| 10,690 | 10,620 | 70 | XO-2Nb (K0) | Griffith (2014)[&&] Figure 9, red line |
| See text | See text | ~0 | Wasp-96b (G8) | Nikolov et al. (2018[b])[&&&] |
| See text | See text | ~0 | Wasp-52b (K2) | Louden et al. (2017)[$$$] |

[*]Space Telescope Imaging Spectrograph (STIS) and the Advanced Camera for Surveys (ACS) aboard the HST. Data suggested Rayleigh scattering consistent with high altitude haze.

[+]STIS and ACS data with Rayleigh scattering assuming clouds with grain sizes linearly proportional to pressure.

[^]WFC3 data on the HST. This work concluded that the Rayleigh slope arose below 0.5 microns and that absorption slope for this planet for 400nm<$\lambda$<1000nm was mainly due to unocculted star spots.

[‰]Data from the HST STIS and High speed Imaging Photometer for Occultations (HIPO) instrument aboard the Stratospheric Observatory for Infrared Astronomy (SOFIA) aircraft. Assuming transmission spectra corresponds to N=2 scale heights.

[°]Gemini south telescope with the R400 grism operating from (515-940)nm.

[°°]Gemini north telescope with the R400 grism operating from (520-930)nm.

[°°°]Multi-Object Double Spectrograph (MODS) aboard the Large Binocular Telescope (LBT) from 330-980nm.

[**]Dobbs-Dixon model of Rayleigh scattering fitted to STIS data on HST. Data suggested weak molecular absorption dominated by continuous opacity due to haze or/and dust. Note that such models have difficulty reproducing the STIS outlier point at 0.95 microns which, if real, would weaken the Rayleigh slope of HD 209458b (see their Figure 14).

[#]STIS data from the HST. Bracketed "()" values show TDs without the outlier datapoint at ~589nm possibly related to strong sodium absorption. This datapoint could not be reproduced by model calculations even when assuming x1000 solar metallicity sodium. Negative values in $TD_{blue-red}$ arose mainly due to strong absorption around 970nm by an unidentified absorber. Goyal et al. (2018) (see their Figure 6) however suggested a modest positive slope with $TD_{blue-red}$ ~several tens of ppm.

[##]Combined STIS G430L (where G=(CCD) Grating, 43=#spectra collected, 0L=low resolution) and G750L datasets. The uncertain, strongly absorbing outlier at ~760nm was removed from the analysis. The red interval extended only up to 970nm.

[^^]Inamori-Magellan Areal Camera and Spectrograph (IMACS) from 470-860nm on the Baade telescope at Las Campañas.

[###]FOcal Reducer and Spectrograph (FORS2) on the Very Large Telescope (VLT) and HST STIS data from 360-850nm. Goyal et al. (2018) (see their Figure 6) however suggested a modest slope with $TD_{blue-red}$ ~several tens of ppm. See also Fischer et al. (2016) who suggested cloud-free conditions with a clear sodium detection for WASP-39b using HST STIS. Wakeford et al. (2018) suggested a rather weak absorption slope for the PLATO FC intervals.

[´]Observed with the HST STIS G430L grating based on best-fit model including Rayleigh scattering haze, a grey high-altitude cloud deck and non-pressure broadened sodium and potassium features.

[++]Whereas Bento et al. (2014) and Sedaghati et al. (2016) did not suggest a Rayleigh-absorption slope for WASP-17b, Sing et al. (2016) and references therein and Goyal et al. (2018) (see their Figure 6) suggested a modest slope with $TD_{blue-red}$ ~several tens of ppm.

[%%]Using mainly photometry imaging via the wide-field imager WiFSIP on the STELLA (STELLar Activity) robotic telescope on Tenerife.

[$]Optical System for Imaging and low-Intermediate-Resolution Integrated Spectroscopy (OSIRIS) on the Gran Telescopio Gran Canarias (GTC). "Red" interval extended only up to 925nm. Non-bracketed values are derived from theory including Rayleigh scattering, flux contamination and unocculted spots (continuous black line, their Figure 15) which under-estimates the observed Rayleigh slope (shown in brackets). The origin of this under-estimate (also noted by Deming et al., 2013) is not well understood. Pinhas et al. (2018) discussed challenges due to stellar properties when retrieving atmospheric properties for a range of HJs.

[%%%]Combined literature data taken from e.g. the GTC, the Calar Alto Observatory and STELLA. This study suggested that the Parviainen et al. study discussed above were over-estimated due to starspot contamination and recommended longer-term monitoring.

[$$]Optical ACAM (Auxiliary port CAMera) on the William Herschel telescope. "Red" interval extended only up to 925nm. Rayleigh slope model fitted to $T_{eq}$=852K for transit light curves fitted with a Gaussian process.

[&]Gemini Multi-Object Spectrometers (GMOS) using the B600 and R150 gratings. Transmission spectrum dominated by uniform opacity with no Rayleigh slope, consistent with ~1 micron grain sizes.

[&&]Model fit assuming cloud-free conditions with atmospheric composition fitted to eclipse data (their Figure 3) and a theoretical Rayleigh slope consistent with optical photometry in the U-band and B-band from the Kuiper telescope.

[&&&]FORS2 observations indicated a relatively clear atmosphere with sodium and potassium bands which weakened the Rayleigh extinction gradient in the PLATO FC intervals.

[$$$]obtained using the auxiliary port camera from 400 to 875nm on the William Herschel telescope. Flat optical slope proposed possibly due to silicate cloud layer.

In Table 3, the $\Delta TD_{blue-red}$ values cover a wide range, from negative values (due to e.g. water absorption in the longer wavelengths), ranging through $TD_{blue-red}$ values of ~zero (e.g. due to the presence of particles larger than ~1 micron which weaken the Rayleigh slope), up to high $\Delta TD_{blue-red}$ values of >1000ppm e.g due to strong Rayleigh scattering in clear atmospheres or in atmospheres with fine, scattering hazes. About one third of the cases in Table 3 have a ~zero value for $TD_{blue-red}$ which impacts the HJ statistics of what PLATO can measure. Note that Bétrémieux (2016) (see also García Muñoz et al., 2012) suggested that refraction could also flatten the Rayleigh slope especially at the larger optical wavelengths from 700 to 800nm. Also, note that forward scattering may affect the transmission radius by up to one scale height depending on wavelength (García Muñoz and Cabrera, 2018).

The PLATO FCs could be used to clarify uncertainty in Rayleigh slopes in Table 3 by revisiting known targets such as e.g. for HD 189733b where observed slopes differ by up ~200ppm - possibly due to contamination of the lightcurve by a stellar spot (Table 3). Such differences could be resolved with the PLATO FCs on averaging 10-15 transits. The relatively long planned observing time of PLATO will therefore likely play an important role in mitigating stellar effects. Based on Table 3, for the noise analysis in our work we will adopt two $\Delta TD_{blue-red}$ values, namely 50ppm, a value in the lower range as observed for HD 209458b and 500ppm, a value in the upper range.

A brief word on white light TDs - these are used to deliver basic information on a range of parameters including planetary size, impact parameter, inclination and limb darkening. It is informative to compare red and blue TDs in Table 3 with the observed white light TD e.g. equal to 1.46% (for HD 209458b) (Tsiaras et al., 2016) with corresponding values of 3.46% and 1.30% for K5 and F5 scenarios respectively.

Appendix A2 (Figure A2, sub-panels (a-d)) shows observed $TD_{blue-red}$ for four representative studies cited in Table 3 [note that for some studies shown in this Table, data was either not available in tabular form or could not be read from the relevant Figure in the literature due to datapoint overlap]. Thick blue (red) lines in Figure A2 show the arithmetic mean TD (ppm) values in the blue (red) interval. Shown in bold in each sub-panel title are the arithmetic mean of $TD_{blue-red}$ and its $\pm 1\sigma_{blue-red}$ error where $\sigma_{blue-red} = \sqrt{[e_{blue}^2 + e_{red}^2]}$ and $e_{blue}$, $e_{red}$ denote the mean error over the blue and red wavelength intervals respectively which are calculated via: $e_a/\sqrt{n}$ where '$e_a$' is the arithmetic mean of the error of the datapoints in the red or blue interval and 'n' is the number of datapoints in the interval.

Numerous processes could contribute to the range of $TD_{blue-red}$ (ppm) values as shown in Appendix A2, Figure A2, (a-d). In the blue interval, strong absorption could indicate a clear atmosphere with Rayleigh absorption, but could also indicate strong sodium absorption at 589nm. In the red interval, strong absorption could occur due to potassium absorption at around 768nm but could also indicate strong water bands in the near-IR. Interfering processes such as the presence of stellar plages and stellar limb darkening could also influence the final absorption values derived in both blue and red intervals if such processes are not adequately accounted for in the analysis. The presence of clouds and hazes generally lead to a flattening of absorption features with wavelength depending upon the cloud properties and location. Breaking the above degeneracies requires additional data which can be obtained e.g. from output of atmospheric models constrained by gathering data on e.g. stellar metallicity, climate (from e.g. eclipse and reflection spectroscopy) and clouds, as discussed e.g. in Barstow et al., (2014). Figure A2 in Appendix A2 illustrates the importance of obtaining more data to better constrain transit depth differences, but also suggests that our chosen range of $TD_{blue-red}$ for the noise analysis with PLATO in the section which follow Table 3, is reasonable for our purposes.

**Measurement Uncertainties** - PLATO will determine planetary mass to within 5% and planetary radius to within 3% (see e.g. Marchiori et al., 2019; Laubier et al., 2017) for a statistically significant number of telluric planets in the habitable zone. This corresponds to a noise threshold of ~34 ppm over one hour for ~20,000 stars at stellar magnitudes <11. The non-photonic noise requirement is ~three times lower than photon noise in the frequency range (0.02-10 mHz) (Laubier et al., 2017). Fluxes are calibrated against a standard 6000K G0V PLATO target (V=11) reference star as described in Marchiori et al. (2019). Substituting e.g. from Figure A2(c) (see appendix A2), a signal, S=40ppm (HAT-32b) assuming the standard case noise, N = 34 ppm/hour suggests (S/N)=[40/34]~1.2 for one transit. Increasing the number of transits ($N_t$) to $N_t$=10 decreases the noise by a factor $\sqrt{10}$ hence increases the (S/N) to ~3.7. The conclusion section discusses how much transit data (number of transits, phase curves etc.) are required to detect a range of assumed atmospheres with different $TD_{blue-red}$ signals.

### 6.1.3 Occultation (secondary eclipse)

The main aim here is to assess how well the PLATO filters constrains occultation transits of e.g. hot Jupiters. The main observable is the difference $TD_{blue-red}$ for light emitted and reflected by the planet. To estimate the emitted component, we assume a planetary emission temperature then apply the Planck equation to calculate electromagnetic fluxes in the blue and red intervals. To estimate the reflected component we assume a wavelength-dependent planetary albedo from the literature and multiply by the electromagnetic flux calculated from the Planck equation for the blue and red intervals assuming a stellar temperature. Our main aim is to obtain a first-order estimate of the range of signals and their statistical significance which could be measured by the PLATO filters.

When interpreting actual data, however, one obtains in practice a given measurement of $TD_{blue-red}$ which requires further analysis e.g. to address the degeneracies involved in order to constrain atmospheric properties. Typically, to achieve this one would apply atmospheric models constrained by additional data. For example, model variables such as atmospheric composition, circulation, planetary albedo and cloud properties are constrained by additional observations such as atmospheric scale height, (brightness) temperature and phase curve amplitude in the IR and visible. There would follow iterative minimization (for more details see e.g. Barstow et al., 2016[b]) between model output and PLATO $TD_{blue-red}$ data. The result would be an improved constraint of combinations of degenerate variables characterizing atmospheres.

Observed secondary TDs for HJs typically vary from tens to hundreds of ppm e.g. 64ppm for KOI-196b (Santerne et al. 2011) and 620ppm for TrES-2b (Croll et al., 2010) peaking in the IR. Observed HJ geometric albedos vary from a few percent up to several tens of percent (see section 1).

Figure 1 shows calculated HJ (planet/star) flux ratios due to planetary emission and reflection assuming planetary properties of HD 209458b. For reflection we assume the wavelength-dependence of geometric albedo observed via polarimetry (Berdyugina et al., 2011) (appendix A3) for HD 189733b near superior conjunction (observations by Evans et al., 2013) also suggested enhanced reflectivity at shorter wavelengths for this planet but with a somewhat weaker wavelength dependence suggestive of optically thick reflective clouds on the dayside. Then, we determine the (planet/star) flux ratio with planetary reflected starlight from equation 9 assuming the planetary properties ($1.35R_{jupiter}$, $instellation_{HD209458b}$, $T_{brightness}=1316K$) of HD 209458b orbiting G5V, K5V and F5V stars. We assume a medium to high geometric albedo, $A_{planet}$ with a rather strong wavelength-dependence ranging from $A_{planet}$=0.48 at 500nm to $A_{planet}$ =0.01 at 1000nm (see appendix A3) based on available data for HD 189733b. Note however that the wavelength-dependence of HJ albedos is generally not well-known; initial estimates for the observed albedo of HD 209458b suggest a rather low value ($A_{planet}$ ~0.038) (Rowe et al., 2008).

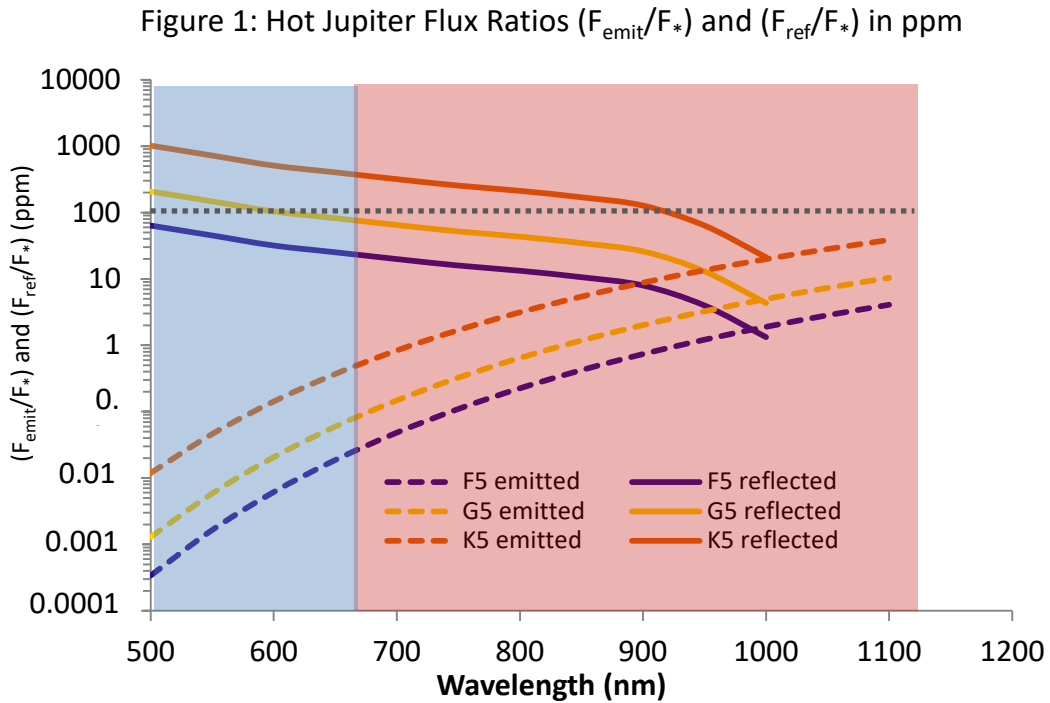

Figure 1: Hot Jupiter Flux Ratios ($F_{emit}$/$F_*$) and ($F_{ref}$/$F_*$) in ppm

Figure 1: Planetary reflection and thermal emission fluxes divided by stellar flux shown as a function of wavelength. The emission calculation assumes black body emission in 50nm intervals for a planet with the properties of HD 209458b assuming fixed $T_{eq}$=1316K taken from the open exoplanet catalogue (Varley, 2015) orbiting F5V, G5V and K5V stars (see Table 1 and text). The reflection calculation is based on equation 8 assuming the albedo in appendix A3. Blue and red-filled areas show the wavelength interval range for the PLATO FCs. Dotted grey horizontal line marks the flux ratio=100 value.

Figure 1 suggests that whereas HJ emission flux ratios are too weak in the blue (dashed lines, <0.01ppm) to be detected by the PLATO FCs, the stronger reflection components (solid lines, several tens up to ~1000ppm) are comparable in range to the primary TDs (see Table 3) and are therefore considered in our S/N analysis (see 7.0). Figure 1 suggests that flux ratios from emission are comparable to reflection in the near IR (see also Table 4 below). The three emission cases (dashed lines) in Figure 1 weakly converge together on going from 500nm to ~1000nm (by ~a factor 3 in the flux ratios shown) - this effect is mostly due to wavelength-dependent changes in the stellar spectra although such effects appear only weakly when considering the large and logarithmic y-axis range. Although Figure 1 provides a good first indication, HJs can of course deviate from the black bodies assumed there e.g. due to clouds (see e.g. Shpor and Hu, 2015). Figure 1 suggests that the emission ratios shown for cooler, smaller K-stars are increased due to a decreased stellar flux for the smaller, cooler K-stars. The wavelength dependence of the reflected fluxes in Figure 1 mainly arose to the assumed dependence of geometric albedo with wavelength as shown in appendix A3.

Table 4 shows values (ppm) from Figure 1 averaged over the PLATO FC red and blue intervals. Hereafter we refer to the flux ratios shown in Figure 1 as the transit depth signals arising due to planetary emission and reflection which are required to be resolved in the S/N analysis.

Table 4: As for Table 3 but for flux ratios (ppm) at occultation calculated from Figure 1.

| Star | TD$_{blue}$ (ppm) Emission Reflection Total | TD$_{red}$ (ppm) Emission Reflection Total | TD$_{blue-red}$ (ppm) Emission Reflection Total |
|---|---|---|---|
| F5 | 0.007 41.47 41.47 | 1.28 8.11 9.39 | -1.28 33.36 32.08 |
| G5 | 0.02 135.35 135.37 | 3.36 26.47 29.93 | -3.34 108.88 105.54 |
| K5 | 0.14 663.45 663.59 | 13.34 129.74 143.08 | -13.19 533.71 520.51 |

Table 4 shows that HJ reflection (varying from tens to hundreds of ppm) is ~3-4 orders of magnitude larger than emission in the PLATO FC blue interval and ~1 order of magnitude larger in the red interval.

### 6.2 Ultra Hot Jupiters

### 6.2.1 Primary Transit

Table 5 is as for Table 3 but for Ultra Hot Jupiters.

Table 5: Primary transit depths (ppm) averaged over the PLATO FC wavelength intervals blue (500-675nm, TD$_{blue}$), red (675-1125nm, TD$_{red}$) (unless stated otherwise) and their difference (TD$_{blue-red}$) for Ultra Hot Jupiters taken from the literature.

| TD$_{blue}$ (ppm) | TD$_{red}$ (ppm) | TD$_{blue-red}$ (ppm) | UHJ (stellar class) | Reference |
|---|---|---|---|---|
| 14, 472 | 14,376 | 96 | WASP-12b (G0) | Sing et al. (2013)[*] Figure 13, purple line |
| 14,500 | 14,300 | 200 | WASP-12b | Stevenson et al. (2014)[+] Figure 21, dotted line |
| 19,700 | 19,700[$] | ~0[$] | WASP-19b | Huitson et al. (2013)[$] |

| | | | (G8V) | |
|---|---|---|---|---|
| 20,200 | 20,200[**] | ~0[**] | WASP-19b | Sedaghati et al. (2015)[**] (and references therein) |
| 19,400 | 19,400[^] | ~0[^] | WASP-19b | Espinoza et al. (2019)[^] Figure 13, black solid squares |
| 25,800 | 25,800[++] | ~0[++] | WASP-43b (K7V) | Murgas et al. (2014)[++] |
| 13,666 | 13,065 | 601 | WASP-103b (F8V) | Lendl et al. (2017)[$$] Figure 7 dashed blue line |
| | | ~0[^^] | KELT-9b (B9.5-A0) | Hoeijmakers et al. (2018) Kitzmann et al. (2018)[^^] |

[*]Rayleigh scattering model assuming a pure $H_2$ atmosphere which yielded a good fit to the datapoints taken with the STIS aboard the HST.

[+]Rayleigh slope fitted to STIS and Gemini GMOS data.

[$] Using STIS G750L data. The four datapoints obtained did not yield a discernable Rayleigh extinction feature.

[**]FOcal disperser and low dispersion Spectrograph 2 (FORS2) on the Very Large Telescope (VLT). No clear Rayleigh extinction feature was evident (see e.g. their Figure A.1). Note that Sedaghati et al. (2017) (their Figure 2) suggested some evidence of a Rayleigh extinction feature but occurring mainly from 0.4 to 0.5 microns i.e. shortward of the PLATO FC range.

[^]Magellan/IMACS combined with HST/STIS data which reported a featureless atmospheric spectrum in the optical/UV and cautioned against stellar faculae as false positives for atmospheric Rayleigh extinction .

[++]Osiris data on the GTC. Difficult to assess Rayleigh extinction signal in the PLATO FC intervals due to lacking wavelength coverage and interference by e.g. sodium absorption.

[$$]Observations from GMOS-North plus optical data using the Gamma-Ray Burst Optical/Near Infrared Detector (GROND) imager on the Danish telescope at the European Southern Observatory (ESO).

[^^]TDs adapted from theoretical transmisssion spectrum KELT-9b; this object at low resolution shows no Rayleigh slope and behaves as a near-perfect black body with negligible albedo, D. Kitzmann, personal communication.

Table 5 suggests that some UHJs have either a weak (or ~zero) Rayleigh extinction (possibly due to clouds) whereas others have large slopes of up to several hundred ppm. Note that the high temperature and low atmospheric weight of UHJs favour extended atmospheric scale heights (equation 3) hence strong spectral absorption features in non-grey UHJ atmospheres. For the UHJ primary transit noise analysis (see 7.0) we assume a strong $\Delta$TD=601ppm based on WASP-103b (Table 5).

**6.2.2 Secondary Transit Depths**

Figure 2 is as for Figure 1 but for UHJs:

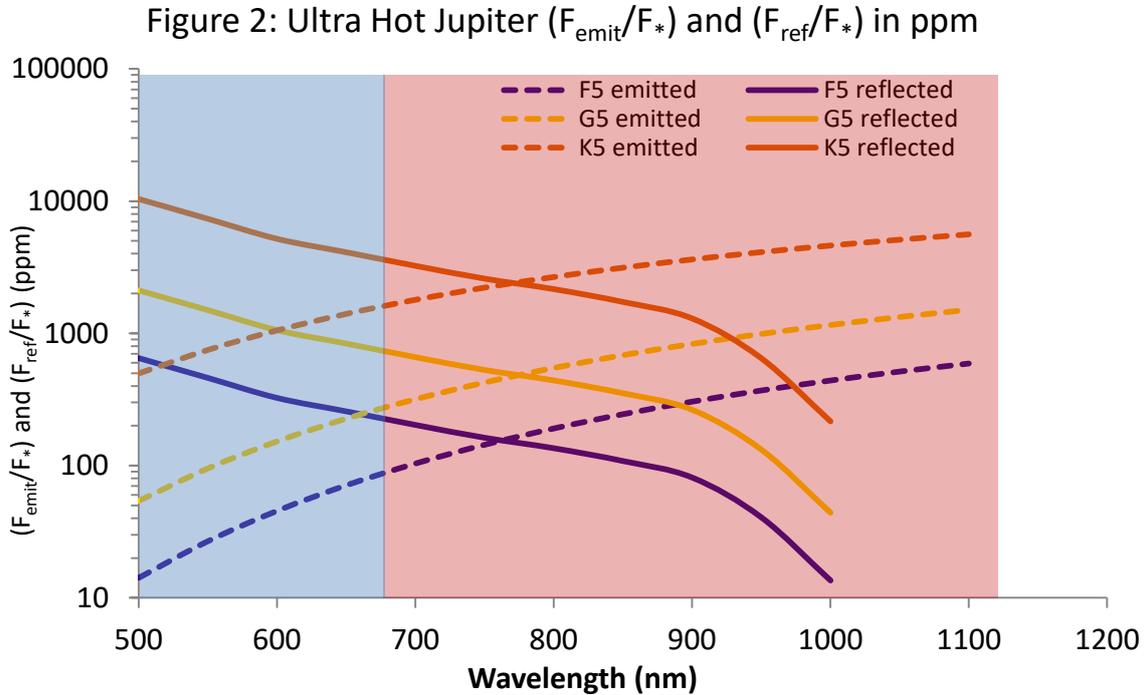

Figure 2: As for Figure 1 but assuming the planetary properties of UHJ WASP-103b ($T_{eq}$=2508K, $R_p$=1.528$r_{Jup}$) and with the same albedo (for HD 189733b) as shown for Figure 1. WASP-103b orbits an F8 (~6300K) star but we assume here a UHJ analogue with similar planetary properties orbiting an F5 (T=6540K) star and then scale to the G5 and K5 cases in order to calculate the same stellar cases as for the HJ analysis in section 6.1.

Figure 2 suggests that UHJ flux ratios (F) increase by several orders of magnitude compared to the HJ case in Figure 1. Some values in Figure 2 are very high (>1000ppm) which could be an artefact of our scaling and our assumption that UHJs exist at such close-in orbits around cooler stars. Unlike Figure 1, F values for UHJs due to emission (dashed lines) and reflection (continuous lines) in Figure 2 have similar magnitudes over the PLATO FC red interval. By comparison, observed values for the UHJ Kelt-9b (Hooten et al., 2018; their Figure 3 and references therein; $T_{Kelt-9b}$ ~4050-4600K) (e.g. Yan and Henning, 2018) suggested somewhat lower values i.e. ($F_p/F_*$)=1000±~80ppm at 850nm (for the sum of planetary emission and reflection) compared with Figure 2. This difference could have arose possibly due to the rather straightforward blackbody assumption in Figure 2 or could be a hint for the presence of clouds on Kelt-9b. The frequency of such objects in nature remains to be established. Table 6 shows mean F values calculated from Figure 2 for the UHJs averaged over the PLATO FC intervals:

Table 6: As for Table 4 but for UHJ (Planet/Star) flux ratios (F) calculated from Figure 2.

| Star | $F_{blue}$ (ppm) Emission Reflection Total | $F_{red}$ (ppm) Emission Reflection Total | $F_{blue-red}$ (ppm) Emission Reflection Total |
|---|---|---|---|
| F5 | 39.4 422.9 462.2 | 61.7 311.2 373.0 | -22.4 111-6 89.3 |
| G5 | 132.1 1376.7 1508.8 | 197.9 1013.2 1211.2 | -65.9 363.4 297.6 |
| K5 | 925.9 6750.0 7675.9 | 1249.9 4968.0 6217.9 | -324.0 1782.0 1458.0 |

Table 6 shows (Planet/Star) flux ratios (F) and suggests that whereas reflection dominates, the UHJ emission component still makes up 10-20% of the and is relatively more important for the red intervals and for the K5 case. Figures in the final column suggest that the value of $FR_{blue-red}$ from emission is negative (i.e. more emission in the red interval) consistent with blackbody behaviour at UHJ temperatures. The reflection component is positive since the implemented dependence of geometric albedo upon wavelength (Appendix A3) peaks at lower wavelengths. Importantly, Table 6 suggests that F values measured for UHJs by the PLATO FCs are degenerate in emission and reflection which will need additional information (e.g. numerical modeling of the planetary brightness temperature) to break.

Note that the blackbody assumption in Figure 2 may under-estimate UHJ emission. For example, the UHJ WASP-33b was suggested to have fluxes are increased by ~50% compared with blackbody fluxes from about 1 micron towards shorter wavelengths (Haynes et al., 2015) possibly due to absorption by TiO (see their Figure 4). That study observed an eclipse depth of 1050ppm at e.g. 1.135 microns (their Table 7) for this UHJ ($R_p$=1.6$R_{jup}$; $T_{day}$~3000K; $T_*$=7400K, $R_{star}$=1.44$R_*$).

### 6.2.3 Phase Curves for UHJs at 10pc

Figures 3a, 3b and 3c shows (Planet/Star) flux contrast ratios over the half-phase curve for WASP-103b analogues placed 10pc away orbiting an F5 star for planetary emission

only (Figure 3a), planetary reflection only (Figure 3b) and the sum of emission plus reflection (Figure 3c) respectively:

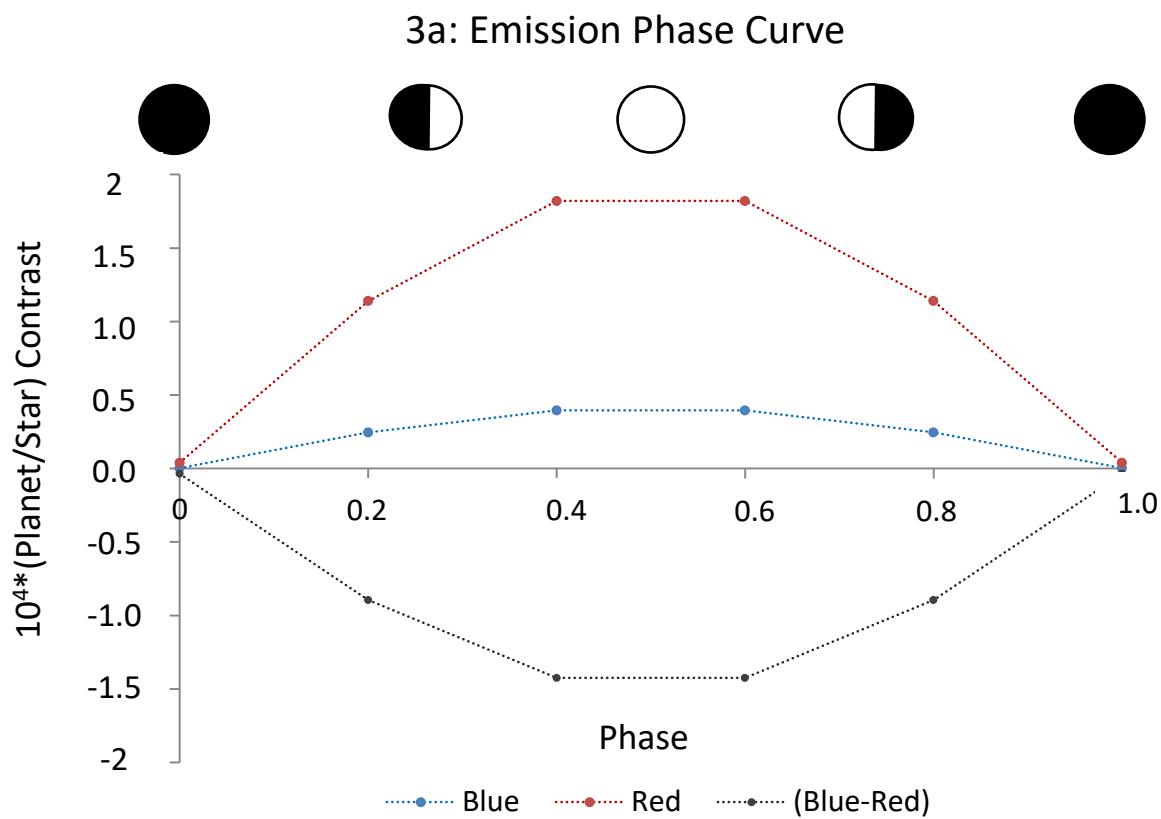

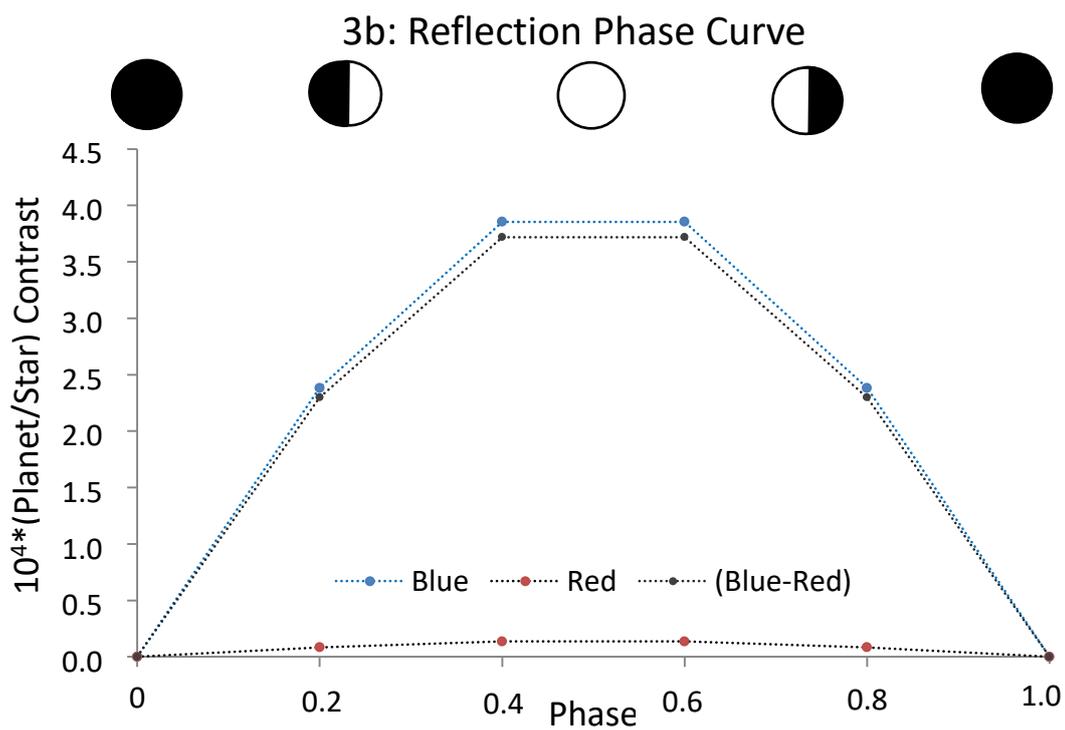

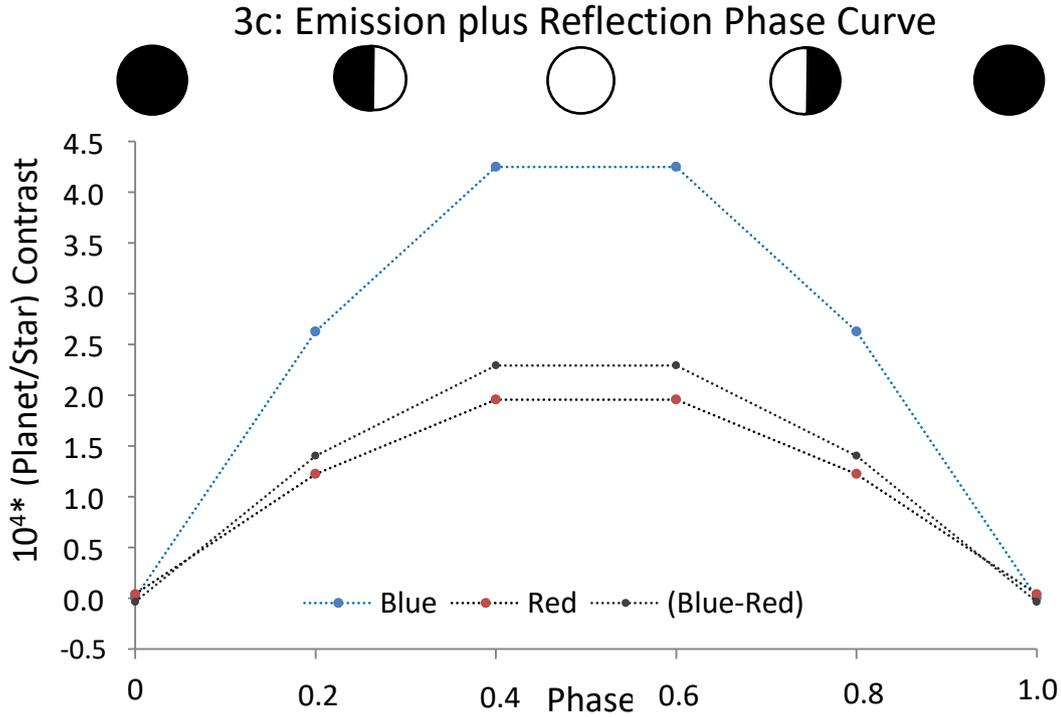

Figure 3: Estimated planetary (planet/star) contrast ratio phase curves sampled by the PLATO FCs allowing for CCD Quantum Efficiencies for an UHJ placed at 10pc assuming the planetary properties of WASP-103b (see Figure 2 legend). Data is shown for 6 equidistant points on the phase curve starting at the nightside during inferior conjunction.

In Figure 3, the six datapoints show the phase curve fluxes calculated via:

$$\text{Flux}_{\text{phase cuve}} = \sin(2\pi n)*\text{Flux}_{\text{FC\_interval\_day}}(\lambda) + (1-\sin(2\pi n))*\text{Flux}_{\text{FC\_interval\_night}}(\lambda) \quad (14)$$

where n=(0,0.1…0.5); interval refers to the FC blue and red wavelength intervals; $T_{\text{day}}$=2508K and $T_{\text{night}}$=1591K (Keating and Cowan, 2018), $R_{\text{planet}}$=1.528$R_{\text{Jup}}$) calculated from the Planck black body equation in 50nm intervals. Figure 3b shows similarly shows planetary reflection from equation (8) assuming an F5 central star with albedo wavelength- dependence shown in appendix A3 scaled at zero phase (nightside) to a value of zero i.e. assuming a nominal value of zero scatter (this assumption may not apply for thick hazy atmospheres such as on Titan; García Muñoz et al., 2017). Figure 3c shows the sum of planetary emission plus reflection.

Comparing the three sub-panels, the phase curves in Figure 3 suggest that reflection dominates over emission for the albedo values assumed here. Can the PLATO FCs detect changes in (b-r) for the phase amplitudes shown? The maximum phase amplitude signal for the F5 case shown in Figure 3 corresponds to a (Planet/Star) flux ratio difference in the blue minus red intervals, ($FR_{b-r}$) at full occultation equal to -22.4 and 111.6ppm (Table 6) for emission and reflection respectively. If we apply the rather conservative conditions from Figure 3 that extracting reasonable information over the phase curves requires resolving ~one order of magnitude lower TDs than the maximum phase (full occultation) TDs shown in Table 6, this would suggest $TD_{(b-r)}$ values of ~-2.2 and ~11.2 ppm (F5) for emission and reflection respectively. For the more favorable G5 and K5 cases (Table 5) these values increase to $FR_{(b-r)}$~-6.6 and ~36.3 ppm (G5) and $FR_{(b-r)}$-32.4 and 178.2 ppm (K5) for emission and reflection respectively (see results section for S/N analysis).

## 6.3 GJ 1214b and GJ 1214b like objects (GLOs)

We first calculate atmospheric spectra for GJ 1214b using two models (6.3.1). Then we compare the range of primary transit depths in the PLATO FCs calculated from these two models and from other studies in the literature (6.3.2) hence choose representative TDs for the S/N analysis which follows.

### 6.3.1 Comparison of GJ 1214b atmospheric spectra using two models

There are two aims here, firstly to compare TDs relevant to the PLATO FC for different assumed atmospheres and secondly to understand differences between theoretical spectra in the literature.

**Model A** developed by D. Kitzmann calculated spectra from atmospheres having chemical concentrations from the FastChem (Stock et al., 2018) chemical equilibrium package assuming an isoprofile temperature of 470K (Miller-Ricci and Fortney, 2010) and with updates for the atmospheric column climate module described in Kitzmann (2016). The same concentrations and temperature profiles were also used as input for spectral model B (see below). Calculations assumed a planet with $R_{planet}$=2.678$R_{earth}$, $M_{planet}$=6.55$M_{earth}$ with 2 bar atmosphere added to this radius. Opacities were calculated using HELIOS-K (Grimm and Heng, 2015) which took as input HITEMP 2010 (for $CO_2$, CO and $H_2O$) and HITRAN (2012) for the remaining species including Rayleigh scattering and collision induced absorption. Water lines were cut at 25cm$^{-1}$ and overlaid with the water continuum calculated using the Mlawer-Tobin-Clough-Kneizys-Davies (MT-CKD) approach and the transmission spectra were normalized from 0.7 to 1.0 microns to yield an overall transit depth of 1.35% as observed by the M-Earth transit survey white light filter. Tholin hazes were treated based on properties discussed in Howe and Burrows (2012) with Mie scattering based on Kitzmann and Heng (2018). Water refractive indices were taken from the international water steam tables with a depolarisation rato of 3x10$^{-4}$ and including a weak temperature and water mass density dependence.

**Model B** developed by A. García Muñoz took as input the same chemical concentrations and temperature profile as model A. Spectral line data were taken from HITRAN2012. Refractive steam indices came from Harvey et al. (1998) and Mie scattering from Mischenko et al. (2000). A water continuum was not included. The general methodology to calculate the spectra is described elsewhere (García Muñoz et al., 2012; García Muñoz and Mills, 2012).

**Scenarios** - we performed eight atmospheric scenarios. (1-4) assumed haze-free conditions with compositions of 0.01xsolar, 0.1xsolar, 1xsolar and pure H$_2$O respectively.  Scenarios (5-8) assumed solar metallicity with monodispersed tholin hazes with refractive indices based on Khare (1984) and extending from (10-0.1) mb with a haze radius of 5nm, 10nm, 20nm (having a number density of N=100cm$^{-3}$) in scenarios (5-7) and 1 micron (having a number density of N=0.1cm$^{-3}$) in scenario 8.

**Comparison** - Figures 4a-e show theoretical transmission spectra calculated by model A. Figures 4a-d show scenarios (1-4) (0.01xsolar, 0.1xsolar, 1xsolar and pure H$_2$O atmospheres respectively. Figure 4e shows scenarios (1-4) plotted together for comparison:

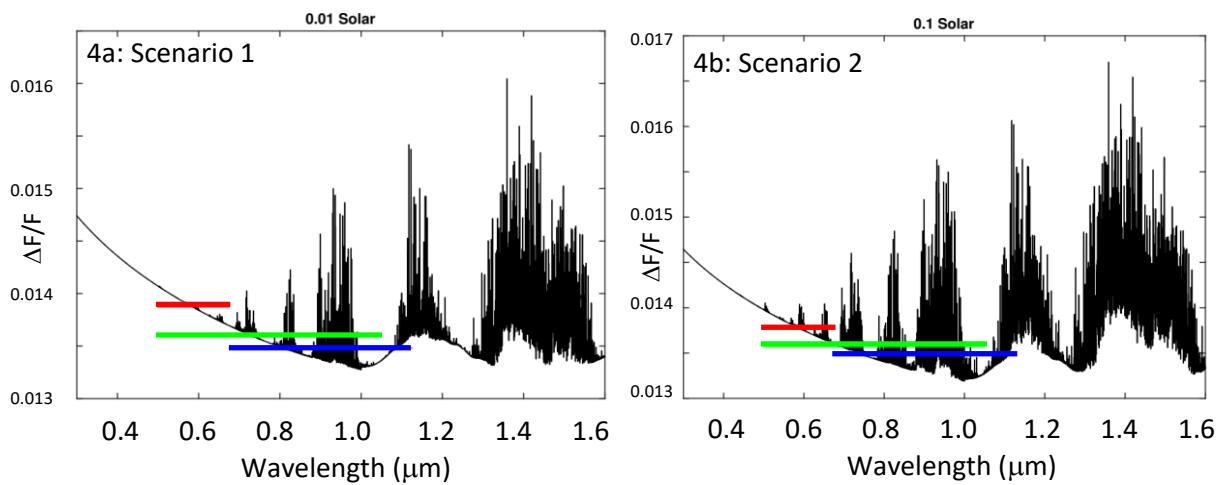

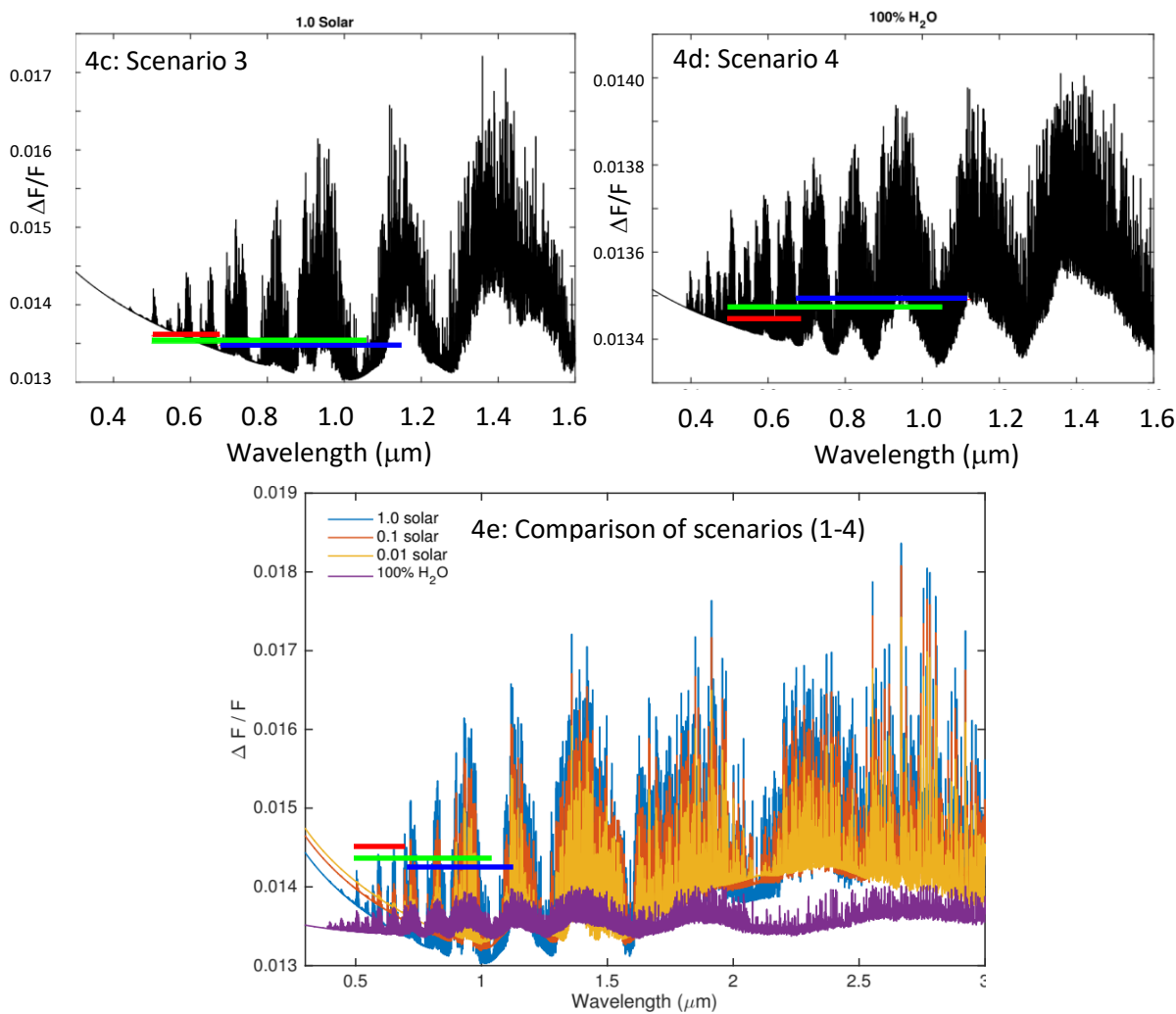

Figure 4: Theoretical transmission spectra for GJ 1214b calculated by model A for scenarios (1-4) (Figures 4(a-d) respectively) and all four scenarios compared (Figure 4e). Note that the wavelength interval includes water band absorption into the IR - beyond the red wavelength interval of the PLATO FCs. Short (long) horizontal blue (red) lines show the PLATO FC blue (red) wavelength intervals respectively. The green horizontal line shows the PLATO "white light" interval (500-1050nm).

The solar case spectrum (Figure 4a) is dominated mainly by $H_2O$, $NH_3$ and $CH_4$ (CO and $CO_2$ concentrations are small and contribute only negligibly). On increasing metallicity (Figures 4a-c) results suggest an increasing contribution of e.g. water bands in the blue FC interval and the dominance of the water bands from the UV to the near IR in Figure 4d. Figure 4e shows the suppression of spectral features for the heavier $H_2O$ atmosphere (continuous purple line).

Figure 5 shows a comparison of GJ 1214b transmission spectra calculated by model A (blue line) and model B (black line) for the 1.0xsolar (Figure 5a) and the 100% water (Figure 5b) scenarios both deconvoluted to a spectral resolution of 4000.

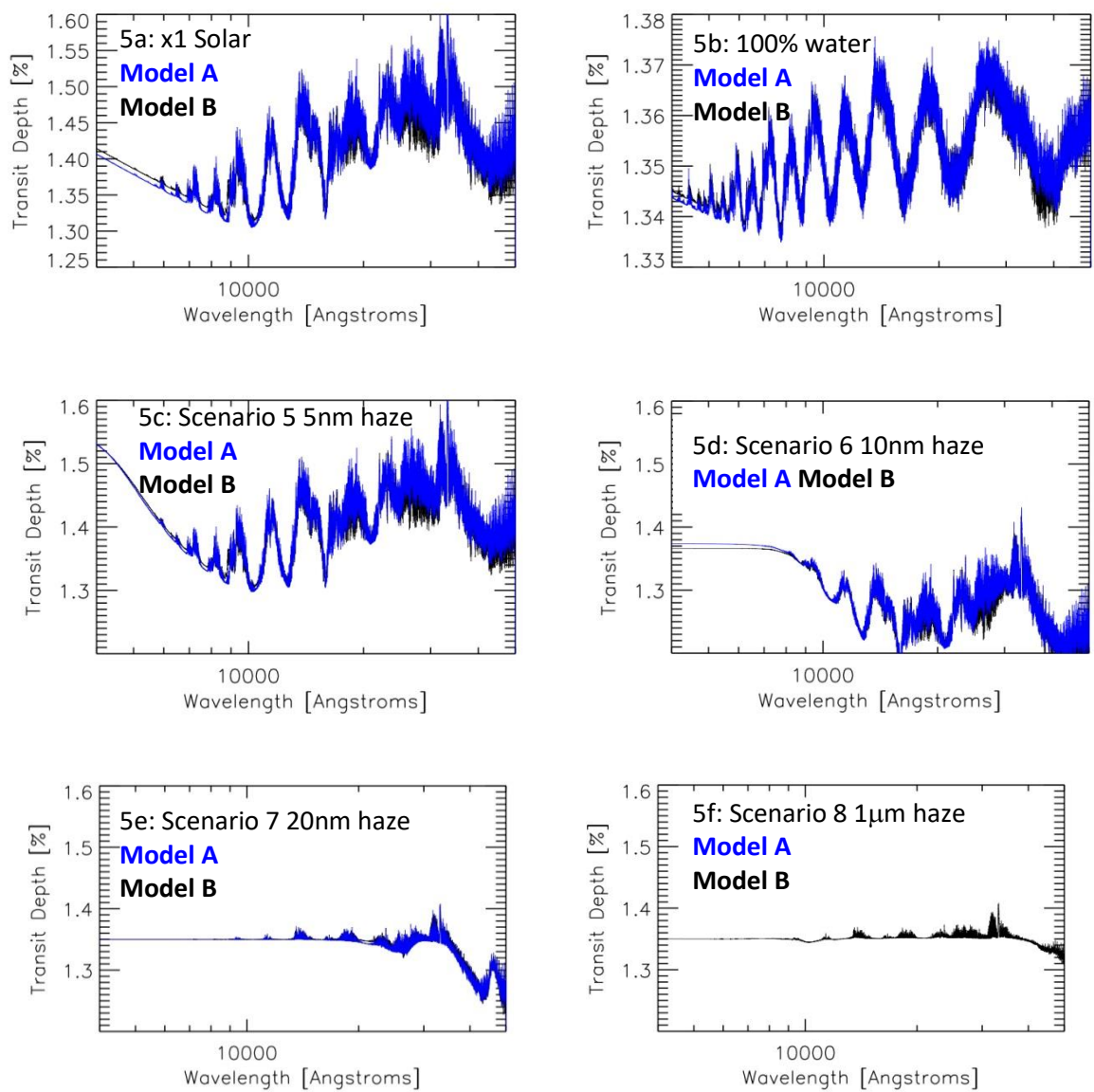

Figure 5: Comparison of GJ 141b transmission spectra calculated by model A (blue) and model B (black) for the x1.0 Solar (Figure 5a) and 100% water (Figure 5b) scenarios and for four scenarios having tholin hazes with radii of 5nm, 10nm 20nm and 1μm (Figures 5c-f).

Figures 5a and 5b suggest that the comparison of model A and model B is overall good. Small differences of around a few percent arise e.g. in the visible region where model A (blue line) calculates slightly lower transit depths and in the near IR where model B (black line) calculates slightly lower transit depths. These arose due to minor differences in e.g. Rayleigh scattering coefficients and gravity treatments in the two models. Figures 5c-f (scenarios 5-8) suggest that scenarios with smaller hazes (radii of 5nm and 10nm) enhance Rayleigh scattering hence the Rayleigh slope in the visible (Figures 5c-d). Larger hazes with radii greater than 20nm however show a flattening in this signal (Figures 5e-f).

**6.3.2 Primary Transit Depths**

For our GLO analysis we assume a hypothetical exoplanet with planetary properties of GJ 1214b as shown in Table 7. For our analysis we place the planet at distances of 10pc and 25pc away from the Earth. We consider only the M-dwarf star case since cases for other stars were at or below the limit of detection with a S/N of x(2-3) lower (not shown).

Table 7: Stellar and planetary properties for GLO scenarios.

| Class | $L_*$ ($L_{sun}$) | $T_*$ (K) | $M_*$ ($M_{sun}$) | $R_*$ ($R_{sun}$) | Magnitude ($M_v$) | $R_{planet}$ ($R_{Earth}$) | $d_p$ (AU) | $P_p$ (d) | Reference |
|---|---|---|---|---|---|---|---|---|---|
| M4.5 | 0.00328 | 026 | 0.157 | 0.211 | 13.05 (10pc) 15.04 (25pc) | 2.678 | 0.0143 | 1.580393 | Charbonneau et al. (2009) |

Table 8 is as for Table 3 but shows an overview of primary TDs (observed and modeled) binned into the PLATO FC wavelength ranges for GLO planets calculated by model A in our work and taken from the literature. Note that values for model B (not shown) are very similar (<1%) to those of model A.

Table 8: As for Table 3 but for GLO primary TDs.

| $TD_{blue}$ (ppm) | $TD_{red}$ (ppm) | $\Delta TD_{(blue-red)}$ (ppm) | GLO Reference(s) |
|---|---|---|---|
| 13,924 | 13,456 | 468 ~0 | GJ 1214b observed Howe and Burrows (2012) e.g. Nascimbeni et al. (2015) |
| 13,873 | 13,483 | 390 398 | GJ 1214b Scenario 1 x0.01 Solar Model A Howe and Burrows (2012) |
| 13,783 | 13,482 | 301 178 | GJ 1214b Scenario 2 x0.1 Solar Model A Howe and Burrows (2012) |
| 13,599 | 13,475 | 124 -250 | GJ 1214b Scenario 3 x1.0 Solar Model A Howe and Burrows (2012) |
| 13,444 | 13,491 | -47 -137 | GJ 1214b Scenario 4 100% Steam Model A Howe and Burrows (2012) |
| 14,100 | 13,500 | 600 | GJ 1214b Scenario 5 x1.0Solar r=5nm haze, Model A |
| 13,700 | 12,950 | 750 | GJ 1214b Scenario 6 x1.0Solar r=10nm haze, (10-0.1mb), Model A |
| 13,500 | 13,500 | ~0 | GJ 1214b Scenario 7 x1.0Solar r=20nm haze, (10-0.1mb), Model A |
| 13,400 | 13,400 | ~0 | GJ 1214b Scenario 8 x1.0Solar r=1μm haze, (10-0.1mb), Model A |
| 14197 | 13379 | 818 | GJ 1214b Howe and Burrows (2012) x1.0 Solar, r=100nm haze (100-10$^{-4}$) bar |
| 13650 | 690 | -40 | GJ 1214b Benneke and Seager (2013) $H_2$, 400ppm $H_2O$, 1micron ZnS haze |
| 13480 | 520 | -40 | GJ 1214b Benneke and Seager (2013) 100% steam |
| 13555 | 540 | 15 | GJ 1214b Benneke and Seager (2013) $H_2$, 400ppm $H_2O$, 0.3micron ZnS haze |
| 6680 | 6840 | -160 | GJ 436b Shabram et al. (2011) x30 Solar |
| 6400 | 5929 | 471 | GJ 3470b observations (PROMPT-8) Awiphan et al. (2016) |
| 3817 | 3817 | ~0 | GJ 1132b (MPG La Silla) Southworth et al. (2017) |
| 385 | 350 | 35 | 55 Cancri e Tsiaras et al. (2016) ($H_2$-He) |
| 365 | 345 | 20 | 55 Cancri e Tsiaras et al. (2016) ($H_2$-He) |

Table 8 suggests that observed GJ 1214b $\Delta$TD values collected around 2012 are rather high (TD=468ppm) suggesting strong Rayleigh extinction (see Howe and Burrows (2012), their Figure 9 and references therein). Their best-fit model to this data was a 1.0xsolar atmosphere having 100nm monodispersed tholin hazes with a number density of 100cm$^{-3}$ extending from 10 to 0.1mb. Subsequent studies however e.g. Nascimbeni et al. (2015) (see also de Mooij et al., 2013 and Cáceres et al., 2014) suggested instead a flat observed spectrum in the optical for GJ 1214b and noted that the 2012 data could have been contaminated with a starspot.

Table 8 suggests that model A (and B) $\Delta$TDs (black) for GJ 1214b for scenarios (1-3) decrease with increasing metallicity due to decreasing scale height hence less extended atmospheres as molecular weight increases. Model A scenario 4 (100% $H_2O$) has a negative TD due to strong absorption by water bands in the near IR. Values marked in red in Table 8 were calculated from the modeling study Howe and Burrows (2012). They assumed $T_{eq}$=470K with chemical equilibrium abundances from Burrows and Sharp (1999) and line lists from Sharp and Burrows (2007). Their treatment of hazes included Mie scattering and assumed complex refraction indices for tholins (see their Figure 2). Comparing values marked in red with those marked in black for GJ 1214b in Table 8 suggests that differences are especially large for the 1xsolar case and the pure $H_2O$ case. This was partly because the earlier line databases (HITRAN 2003) used in the 2012 work (see above) featured considerably fewer water bands in the PLATO FC blue interval. Scenarios (5-8) in Table 8 featured rather large $\Delta TD_{blue-red}$ for the smaller particle cases (scenarios 5,6) due to strong Rayleigh scattering but with a distinct lowering in $\Delta TD_{blue-red}$ for scenarios having particles with larger radii of 20nm and 1000nm (see also Figures 5e, 5f). Whereas model A (values marked in black in Table 8) suggested low $\Delta TD_{blue-red}$ values for GJ 1214b for hazes of 20nm and larger, the Howe and Burrows (2012) study (values marked in red in Table 8) suggested strong Rayleigh extinction ($\Delta TD_{blue-red}$=818ppm) even with relatively large hazes of r=100nm. Data from the Howe and Burrows (2012) study is not considered in our analysis. Blue model values in Table 8 illustrate the challenge of interpreting weak $\Delta TDs$ associated with heavier atmospheres (e.g. Benneke and Seager (2013) considered $H_2O$ and different $CO_2$-$H_2$ mixtures) having degeneracies with hazy atmospheres which will likely require JWST to address (see also Gao and Benneke 2018). $\Delta TDs$ for GJ 436b in Table 8 suggest that heavier atmospheres (30xsolar in this case) can lead to negative values (($\Delta TD_{blue-red}$ =-160ppm) via weak Rayleigh extinction in the blue interval due to low atmospheric scale heights and enhanced absorption by water bands in the red interval. Turner et al. (2016) presented an initial optical transmission spectroscopy data study for several GLOs (and HJs). Improved data is however needed in future to estimate Rayleigh extinction features so these objects are not included in Table 8. For GJ 3470b in Table 8, Awiphan et al. (2016) suggested a light ($H_2$/He) gas envelope with a haze layer in the upper atmosphere. Results of Chen et al. (2017) also tentatively implied a ($H_2$/He) atmosphere but suggested that more data is needed to confirm possible hazes (see also Ehrenreich et al., 2014). Initial results for 55 Cancri e in Table 8 suggest a hydrogen-rich atmosphere.

PLATO white light (500-1050nm) GJ 1214b $\Delta TDs$ from Model A compare quite well with values from Howe and Burrows (2012) (shown in square brackets) for the interval of scenarios (1-4): 13,601 [13,635]; 13,570 [13,574]; 13,504 [13,465] and 13,475 [13,465] respectively. For the GLO noise analysis presented below we assume the $\Delta TDs$ from scenarios (1-8) in Table 8 calculated by model A.

### 6.3.3 Planetary Occultation by Star

The science of occultation observations for determining (Planet/Star) flux ratios of GLOs is still in its infancy. For the hot SE 55 Cancri e (55Cnc e) Tamburo et al. (2018) (see also Demory et al., 2016) suggested a variable observed occultation depth varying from several tens to several hundreds of ppm. Photometric measurements by Dragomir et al. (2012) suggested an upper limit of 0.6 for the albedo of this object. Demory (2014) studied 27

Kepler hot SE candidates using a Markov Chain Monte Carlo method which suggested GLO albedos with values from 0.16 to 0.3 in the Kepler bandpass i.e. higher than those measured for HJs which are in the range from 0.06 to 0.11. Samuel et al. (2014) discussed the albedo of the very hot SEs Corot-7b and Kepler-10b assuming a JWST-type instrument setup. Given the uncertainty, in our work we calculate the (Planet/Star) flux ratio of CoRoT-7b (section 7.0, Table 9) due to reflection assuming the same albedo as for our HJ cases.

## 7. Results

We take as input the observed transit depths from the literature as shown in Table 8. We then calculate the resulting S/N for planetary transmission, emission and reflection assuming noise sources discussed in section 5 for HJs and GLOs orbiting different central stars up to 100pc away. Whereas primary transmission $\Delta$TDs are taken directly from the literature, we calculate for occultation (i) the (Planetary Emission/Star) flux ratios (for the planetary emission component of the occultation) assuming blackbody emission from the Planck formula for planet and star and taking as input the observed brightness temperature, (analogous to the calculation shown in 6.1.3 for HJs) and (ii) the (Planetary Reflection/Star) flux ratio (for the planetary reflection component of the occultation) assuming the wavelength-dependent albedos in appendix A3. To provide an overview of the range of occultation signals which arise, Figure 6 shows calculated photon fluxes (F) (photons cm$^{-2}$ s$^{-1}$) detected by the PLATO FC CCDs allowing for QE and transmission for a range of planets and stars (Figure 6a) placed at 10pc and a zoomed-in selection showing planets only (Figure 6b).

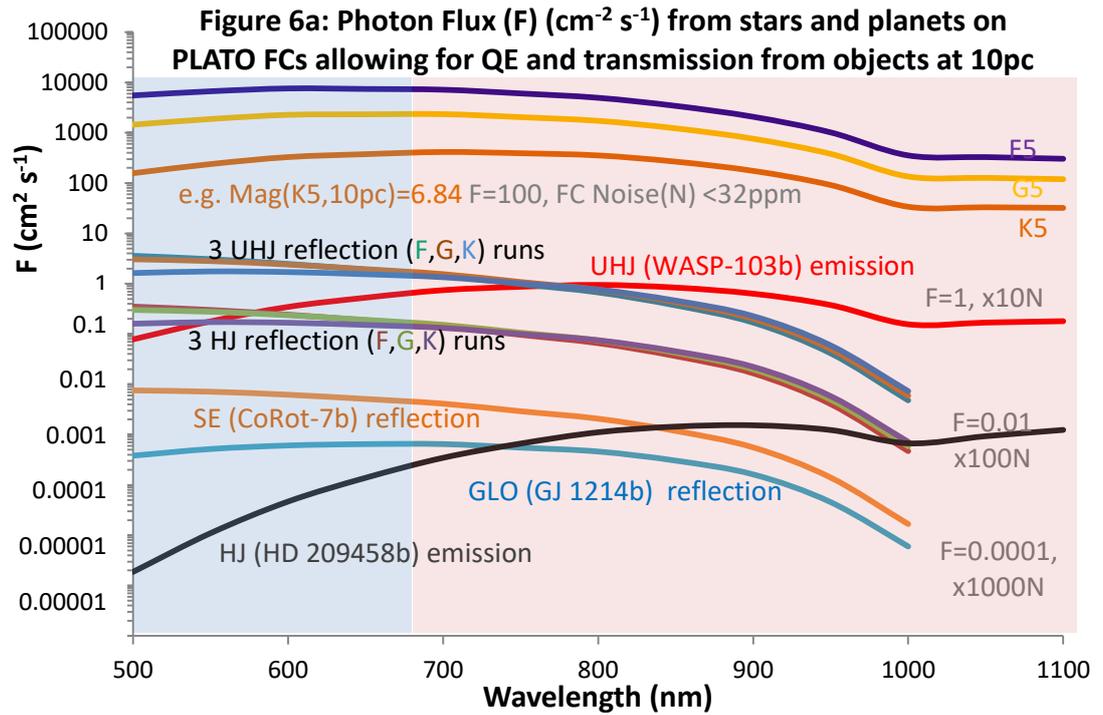

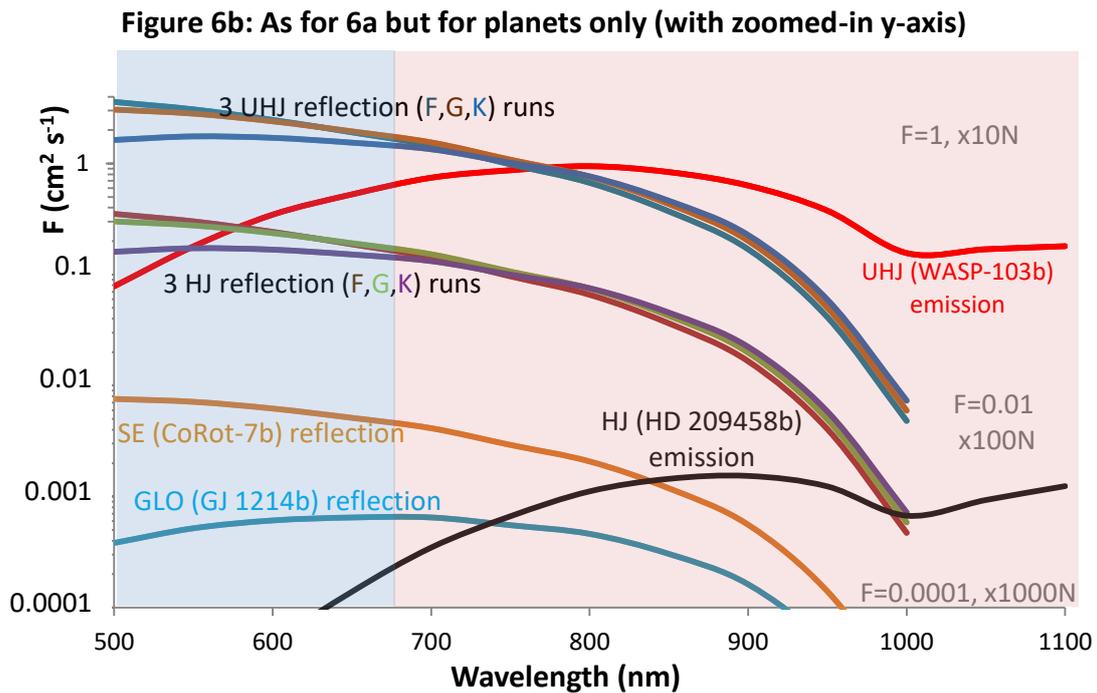

Figure 6: Photon flux (photons cm$^{-2}$ s$^{-1}$) detected by the FC CCDs allowing for QE and transmission for a stars and planets (Figure 6a, upper panel) and planets only (Figure 6b lower panel) assuming black body behaviour calculated in 50nm intervals. Shaded areas show the FC blue and red wavelength intervals. Stellar curves (Figure 6a) are based on data shown in Table 2. UHJ cases assume planetary properties of WASP-103b (R=1.528R$_{jup}$, T=2508K, a=0.01985 AU for the F-star case scaled to have the same instellation for the G and K cases). Geometric albedo values were taken from appendix A3. Reflection fluxes are zero above 1000nm since the assumed albedos are zero in this range. HJ cases assume planetary properties of HD 209458b (section 6). GJ 1214 system properties were taken from Table 7. CoRot-7b (R=1.7R$_{earth}$, a=0.017 AU, T$_*$=5250K, R$_*$=0.82R$_{sun}$; exoplanet.eu). Values shown in grey on the right indicate (from top to bottom) show (i) F (the photon flux, values read directly off the left y-axis) and (ii) the increase in photon noise (PN) as the targets become fainter from top-to-bottom. Changes in PN vary as $F^{-1/2}$ (see equation 10). In Figure 6b, this means that *decreasing* F by e.g. x100, (from a value of e.g. 1 to 0.01 in Figure 6b) leads to an *increase* in PN by $100^{1/2}$ =10 (as shown by grey values on the right of Figure 6b).

Figures 6a and 6b suggests that photon fluxes for the planetary scenarios shown exceed five orders of magnitude in range. In Figure 6 the deviation from Planck behavior shown in the stellar and emission curves redward of 1000nm is related to a low and constant QE. The three (F, G, K-star) UHJ (and HJ) reflection curves cross each other with wavelength due to two effects which oppose each other: on the one hand, the hotter stars emit higher photon fluxes in the blue interval; on the other hand, their planets are placed further away (in order to conserve instellation) where reflection (which scales inversely and quadratically with planet-star distance) is weaker. The reflection fluxes for CoRot-7b are ~one order of magnitude higher than for GJ 1214b e.g. because it receives more radiation and its star is much hotter (5250K cf 3026K for GJ 1214). Figure 6 suggests that for reflection, the (uncertain) dependence of albedo on wavelength could play an important role in determining the difference in photon fluxes between the blue and red intervals. Occultation TDs for reflection of CoRot-7b and GJ 1214b derived from Figure 6 are shown in Table 9:

Table 9: Flux ratio, $F_{ref}$ = (Planet$_{reflection}$/(Star) (ppm) at planetary occultation where the planetary component is for reflection only (no emission) – for CoRot-7b and GJ 1214b taken from Figure 6.

| Object | $F_{ref}$ blue (ppm) | $F_{ref}$ red (ppm) | $\Delta F_{ref}$ (blue-red) (ppm) |
|---|---|---|---|
| CoRot-7b | 2.7 | 0.5 | 2.2 |
| GJ 1214b | 9.3 | 1.6 | 7.7 |

Values in Table 9 are ~(1-2) orders of magnitude lower than (most) TDs discussed above for HJs, UHJs, and GLOs. They are therefore not detectable by the PLATO FCs and are not considered in the noise analysis below. We now present the (S/N) of primary and occultation transit for our various planetary cases orbiting different stars placed at differing distances from Earth.

**7.1 Hot Jupiters**

**7.1.1 Primary Transmission**

**Low $\Delta TD_{blue-red}$ of 50ppm** Figure 7 shows primary transit depth and photon noise for the hot Jupiter (HD 209458b analogue) cases assuming $\Delta TD_{blue-red}$=50ppm (see Table 2 and discussion thereto) for the G-star case with orbital distance scaled to conserve instellation for the K and F cases:

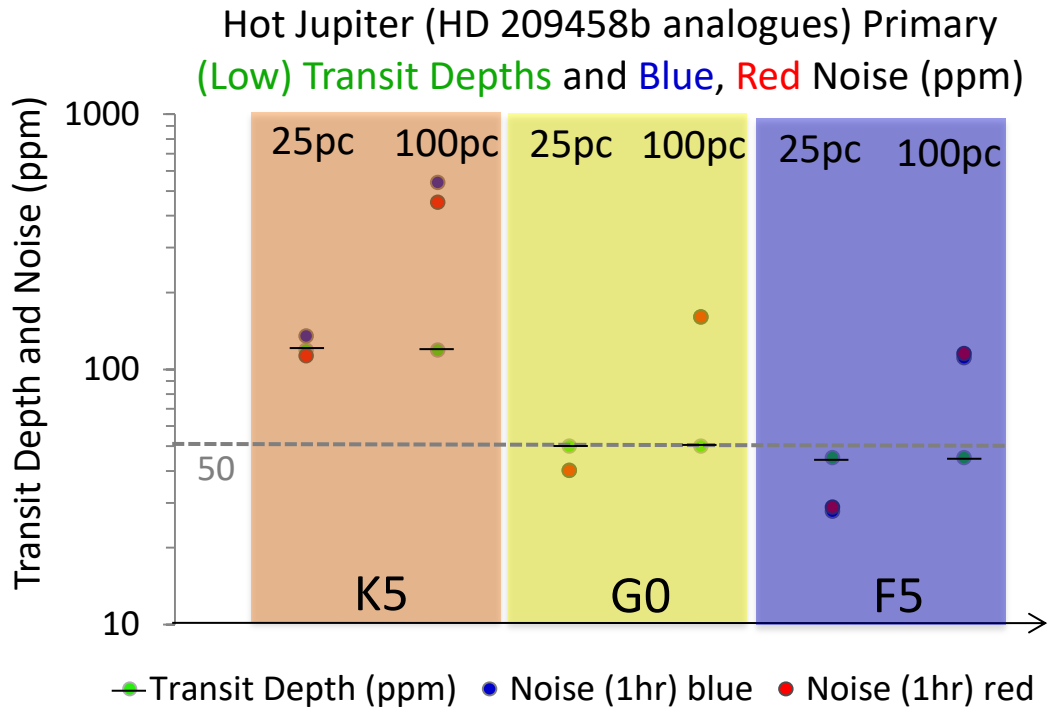

Figure 7: Theoretical Hot Jupiters with planetary properties of HD 209458b orbiting K, G and F-stars placed 25pc and 100pc away from the Earth with magnitudes 12.26, 9.25 (K5); 9.42, 6.41 (G0) and 8.59, 5.58 (F5). Green dots struck through with horizontal black lines show primary transit depth differences for the (blue minus red) FC intervals. These are set to 50ppm (indicated in grey on the y-axis) for the G-star case and then scaled accordingly for the K and F-star cases. One hour noise (ppm) (photon plus instrument) is shown for the blue channel (blue dots) and the red channel (red dots).

Figure 7 suggests that for the G-star case the blue and red noise values overlie (the filters were designed such that each takes 50% of the flux of a G-star as mentioned above). Moving to the smaller K-star scenario leads to an increase in $\Delta TD$ by more than a factor of two and weaker stellar fluxes in the blue spectral region leads to stronger noise in the blue compared to the red channel. For the F-star scenario which has stronger emission in the blue part of the spectrum, the reverse is the case. Figure 8 shows the value: $\sigma(F_{blue}-F_{red})$ calculated from equation (15). This is the signal to noise (S/N) in sigma for the difference measured between the FC blue and red intervals, shown here for the same scenarios as Figure 7 where 'Signal' refers to transit depth (ppm) differences for the $\Delta TD_{blue-red}$ intervals taken from the literature (see e.g. Table 3) and 'noise' refers to the sum of the photon noise plus instrument noise (discussed in 5.3) in the blue and the red intervals:

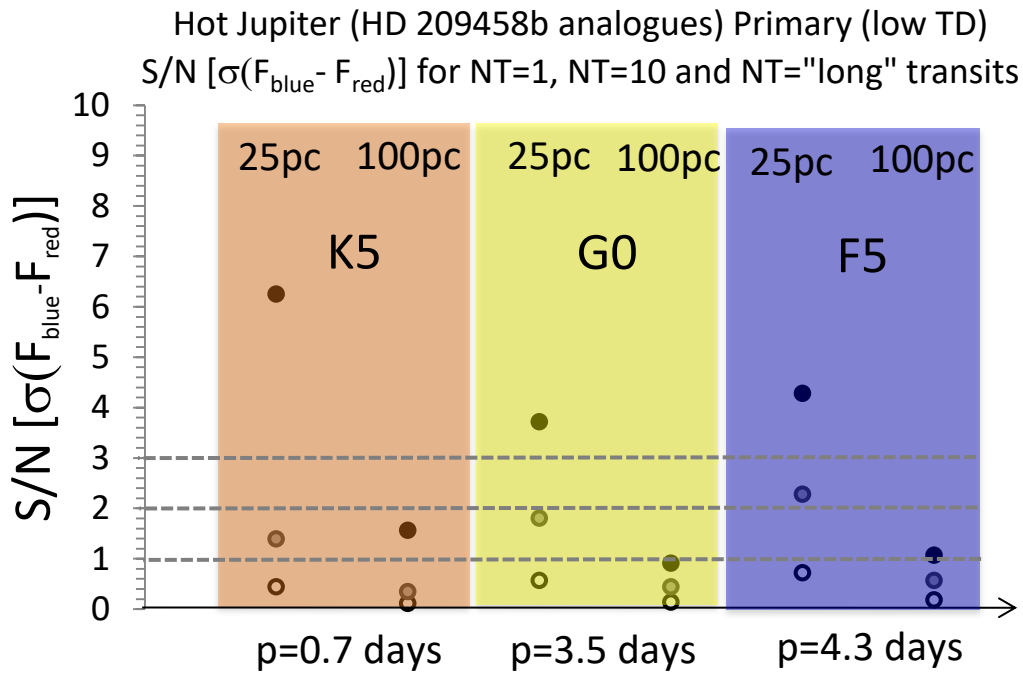

NT=1 (white dots); NT=10 (grey dots); NT=number of transits in 150 days (dark grey dots)

Figure 8: The S/N ($\sigma_{blue-red}$) for the same scenarios as shown in Figure 1. Values are shown for the number of transits, NT=1, NT=10 and NT="long" i.e. averaged over 150 days, assuming $\Delta TD_{blue-red}$=50ppm. Planets are placed around the star where they receive the same instellation as HD 209458b. Planetary periods (p) (days) (shown below the x-axis) are then calculated from Kepler's third law. Grey horizontal dashed lines show S/N=1, 2 and 3.

Figure 8 suggests that HJ signals are detectable at 25pc for an assumed 150-day ("long") pointing for all cases considered but are rather challenging at 100pc. The K-star case has the highest S/N (>6) favored by its short period (P=0.7 days) for the assumed long pointing hence more transits collected during a given observing period. The F-star case has a higher S/N than the G-star case (despite its somewhat longer period) due to suppressed photon noise in the visible (see Figure 1). The 100pc case in Figure 8 suggests values in the range 1<S/N<2 when averaging over the "long" (150 days) period (black dots). For N=10 (grey-filled circles) Figure 8 suggests modest values in the range 1<S/N<2 for the 25pc case and no detection (S/N<1) for the 100pc case.

**High $\Delta TD_{blue-red}$ of 500ppm** Figure 9 is as for Figure 7 but for a HJ transit depth difference in the blue minus red FC intervals equal to 500ppm.

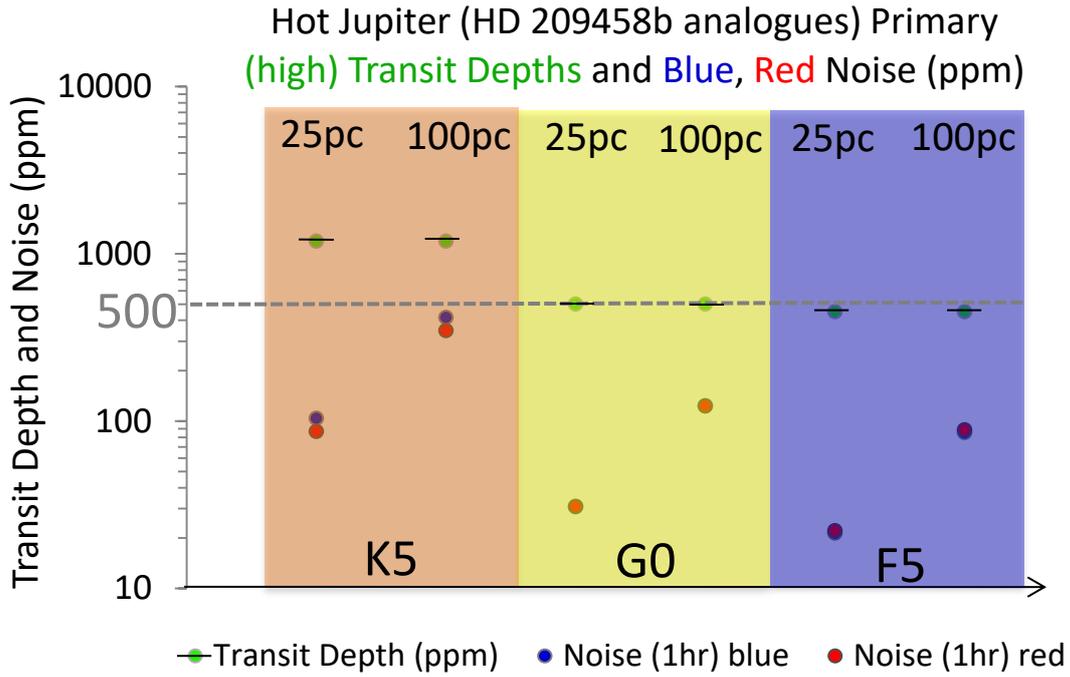

Figure 9: As for Figure 7 but for an assumed high TD difference in the blue minus red FC intervals equal to 500ppm.

Figure 10 shows the S/N ($\sigma_{blue-red}$) for the scenarios shown in Figure

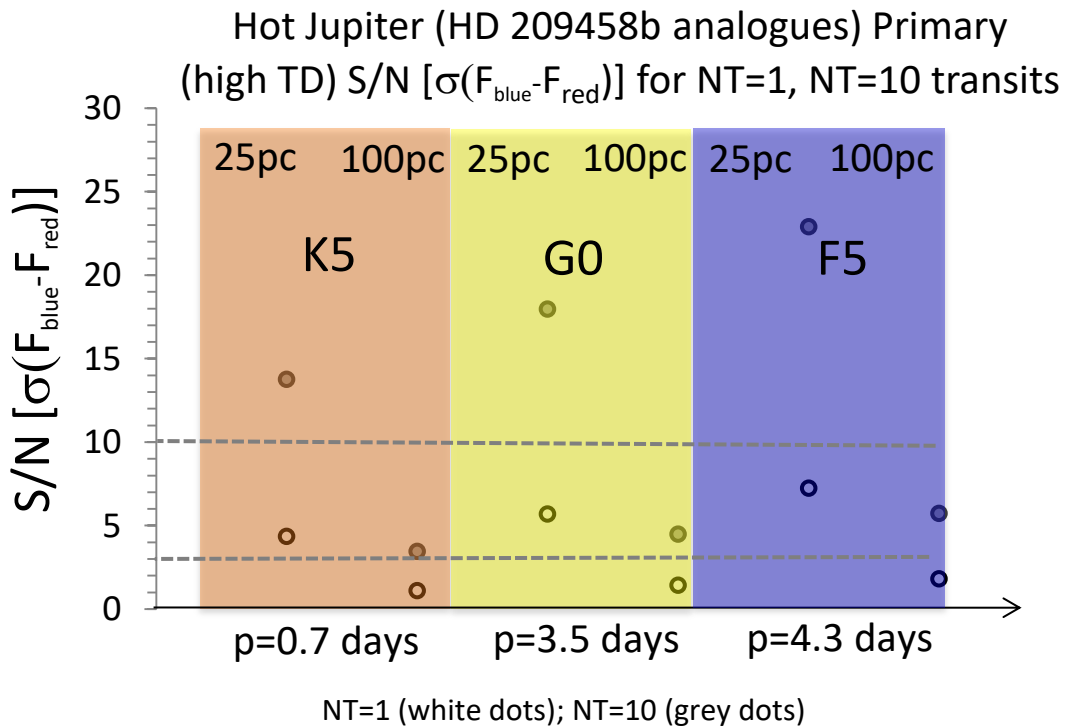

Figure 10: As for Figure 8 but for an assumed high TD difference in the blue minus red FC intervals of $\Delta TD_{blue-red}$ = 500ppm. Grey horizontal dashed lines show S/N=5 and 10.

Figure 10 suggests that unlike the low TD case (Figure 8) signals with a detectability greater than 3σ are now attainable for high TD HJs up to 100pc when averaging over ten transits.

### 7.1.2 Occultation reflection

Note that the HJ emission component (not shown) is too weak (see Table 3 and discussion) to be considered by the PLATO FCs. Figure 11 is as for Figure 9 but for HJ occultation signals based on the (Planet/Star) flux ratios (F) shown in Table 3:

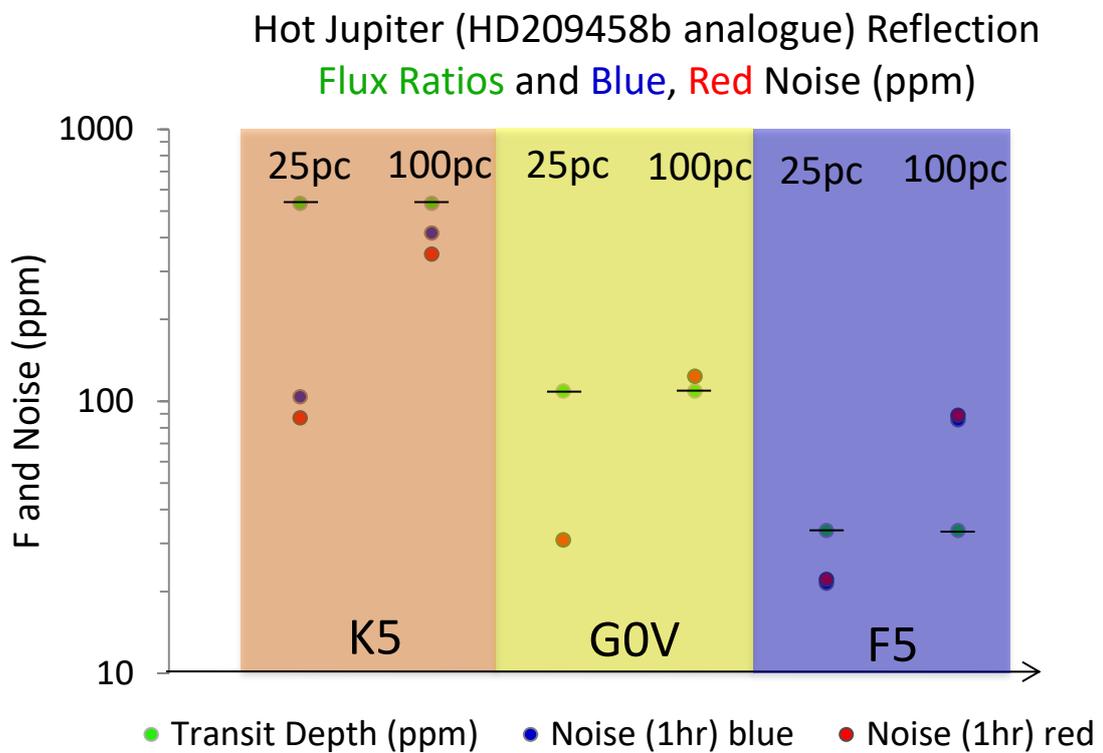

Figure 11: F (=Planet$_{reflection}$/Star) flux ratio from Table 3 and associated FC noise (photon plus instrument).

Figure 12 shows the associated S/N (σ$_{blue-red}$) for the scenarios shown in Figure 11:

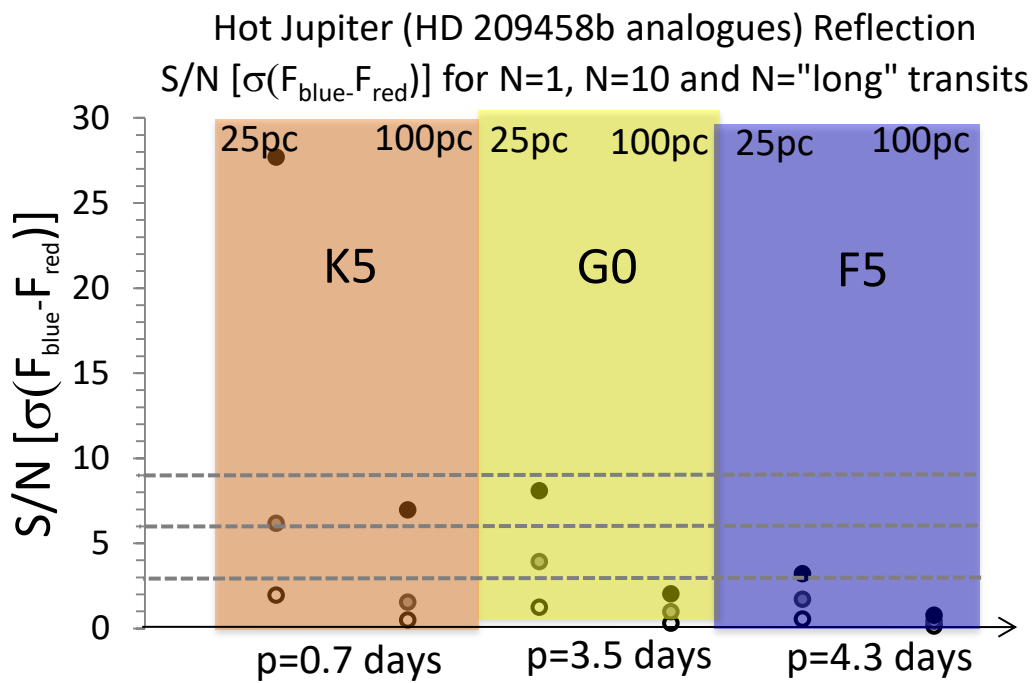

NT=1 (white dots); NT=10 (grey dots); NT=number of transits in 150 days (dark grey dots)

Figure 12: As for Figure 10 but for an assumed high TD=500ppm.
Grey horizontal dashed lines show S/N=3, 6 and 9. "Long"=150 days.

Figure 12 suggests that the K-star and G-star scenarios yield S/N>3σ when averaging over ten transits whereas the F-star scenarios require averaging over the "long" period of 150 days to yield S/N>3σ.

## 7.2 Ultra Hot Jupiters

### 7.2.1 Primary Transmission

Figure 13 shows TDs and noise (photon plus instrument) for the UHJ WASP-103b analogue (F,G,K) star cases assuming a transit depth difference for blue minus red equal to 601ppm from Table 4:

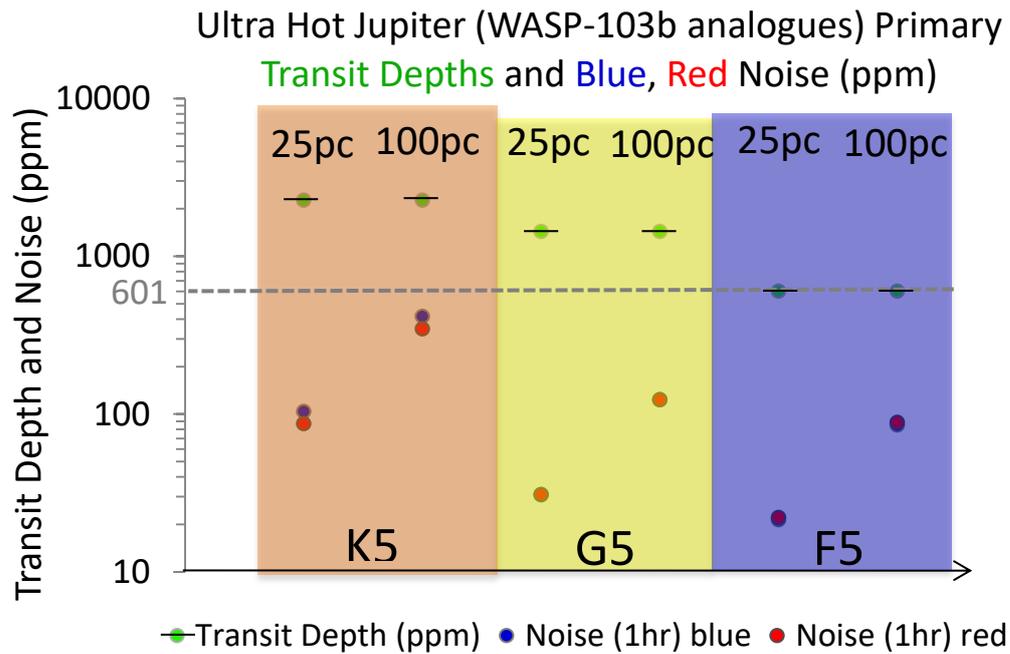

Figure 13: Primary TDs and FC noise (photon plus instrument) for UHJ WASP-103b analogues assuming $\Delta TD_{blue-red}$=601 from Table 5 for F-case scenario indicated by horizontal grey dashed line.

Figure 13 shows high TD differences of up to several thousand ppm larger for the K5 scenarios due to the smaller stellar radius. Figure 14 shows the resulting S/N based on Figure 13:

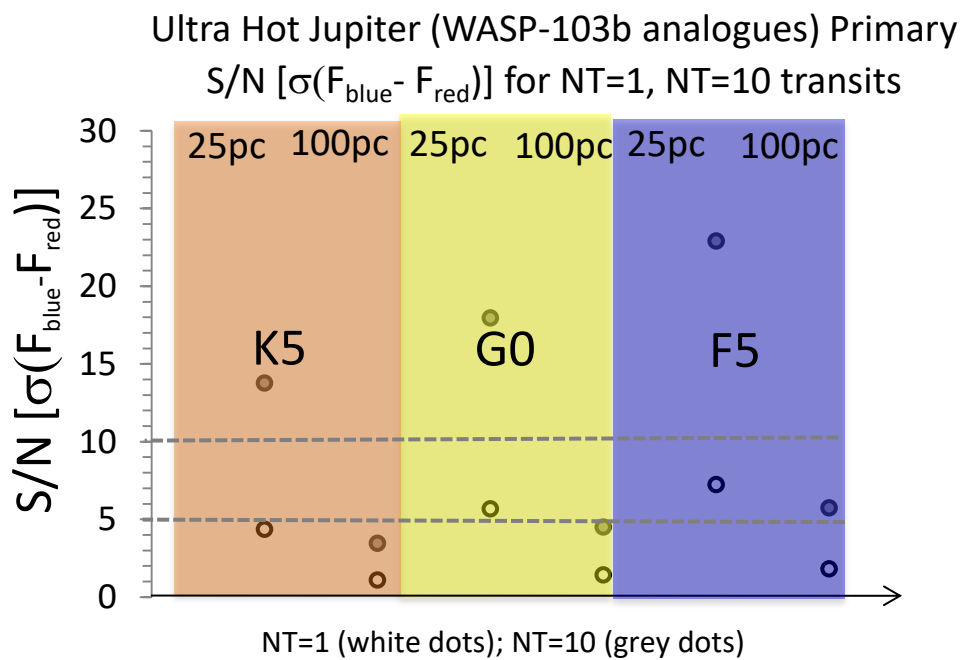

Figure 14: S/N for UHJ primary TD scenarios based on Figure 13. Horizontal grey dashed lines show the five and ten sigma levels.

Figure 14 suggests that a potentially large number of UHJ Rayleigh extinction signals for UHJs could be detectable with S/N>3σ up to at least 100pc when averaging over 10 transits. The quickly orbiting (p<0.5 days) hypothetical K5 case in Figure 14 orbits faster than the fastest known orbiting exoplanets (terrestrial-sized planets orbiting M-dwarf stars, see e.g. Smith et al., 2018) so it may be unrealistic.

### 7.2.2 Occultation

Figure 15 shows differences in (Planet/Star) flux ratios for blue minus red intervals and the noise (photon plus instrument) for the sum of the emission plus reflection fluxes at UHJ occultation:

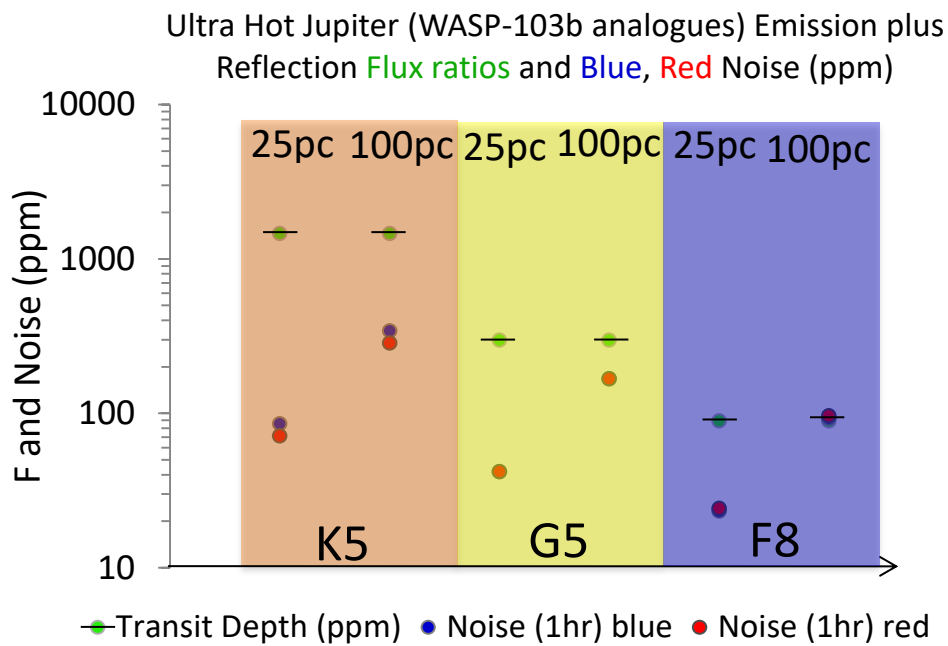

Figure 15: Occultation (emission plus reflection) (Planet/Stellar) flux ratios (F) and FC red and blue interval noise (photon plus instrument) for UHJ WASP-103b analogues assuming flux ratios from Table 5.

Figure 16 shows the UHJ S/N corresponding to Figure 15:

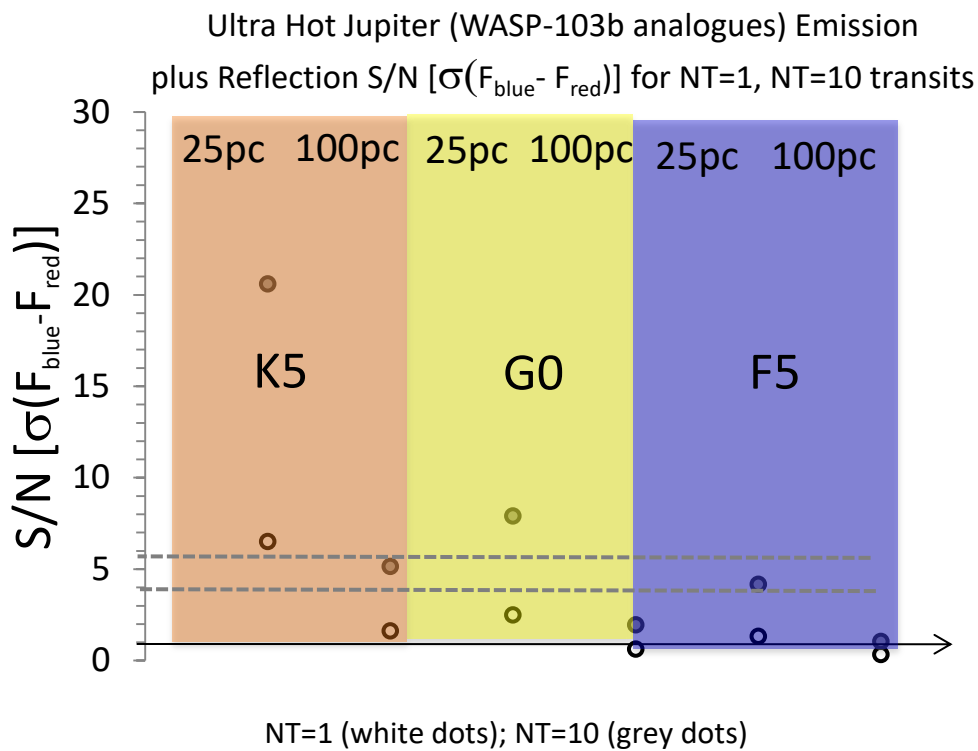

Figure 16: S/N for UHJ primary TD scenarios based on Figure 15.
Horizontal grey dashed lines show the three and five sigma levels.

Figure 16 suggests detection with S/N>3σ for all three cases at 25pc and also for 100pc for the K5 case. As stated in section 6, further information e.g. radiative transfer modeling is required to separate out the emission plus reflection degeneracy observed.

### 7.2.3 Phase Curves

As discussed in section 6, to assess whether phase curve can be extracted by the PLATO FCs we consider TDs reduced by x10 compared with those in Figure 15 assuming the same noise levels. Figure 17 shows the resulting S/N for the three UHJ scenarios as shown in Figure 15:

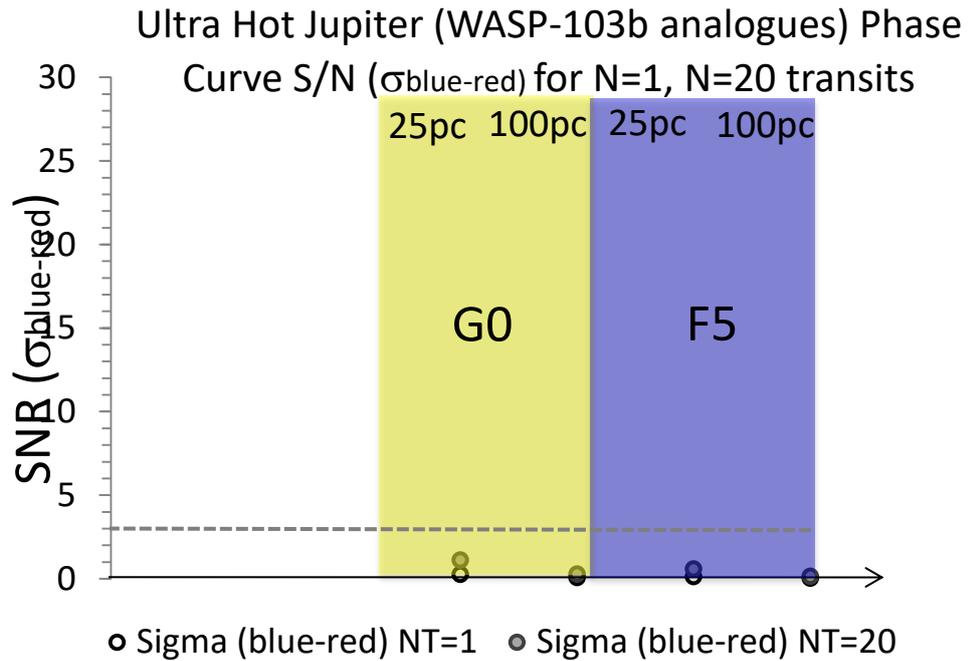

Figure 17: S/N for UHJ primary TD scenarios based on Figure 15. Horizontal grey dashed lines show the three sigma level.

Figure 17 suggests that although detecting UHJ phase curves will be challenging for the PLATO FCs, it could nevertheless be feasible for closer UHJs up to 25pc orbiting e.g. G-stars. Note that the K-star case was not considered in Figure 17 since in order for the planet in this scenario to have a similar insolation as WASP-103b would place it inside the Roche lobe of the star (see e.g. Rappaport et al., 2013).

**7.3 Low Mass Low Density Planets**

Figure 18 shows S/N for the primary $\Delta TD_{blue-red}$ of hypothetical GLO objects with the planetary properties of GJ 1214b orbiting an M4.5 star placed at 10pc and 25pc assuming six different atmospheric compositions as discussed in Howe and Burrows (2012). TDs in Figure 18 are the "Model A" cases shown in Table 8:

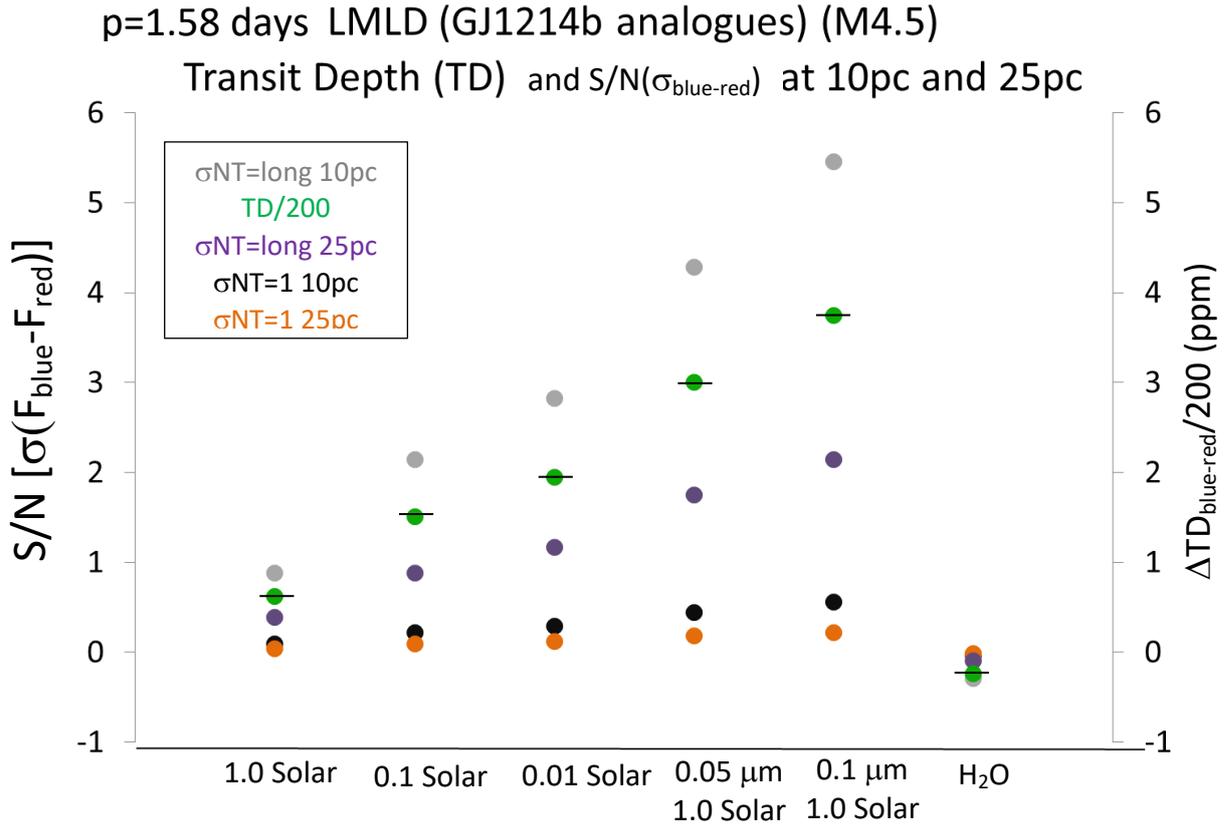

Figure 18: Primary $\Delta TD_{blue-red}$ FC intervals (from Table 8, model A results) and S/N for hypothetical GLO planets with the same radius and incoming installation as GJ 1214b. Planets are placed 25pc and 10pc away and feature six assumed atmospheric compositions which are (1) 1.0xSolar (2) 0.1xSolar(3) 0.01xSolar (4) 1xSolar with 0.05 micron radius tholin haze extending from (10-0.1)mb and constant number density, N=100cm$^{-3}$ (5) as (4) but with doubled radius, and (6) steam atmospheres. Results show S/N for detecting the difference in the PLATO (blue-red) intervals for the number of transits, NT=1 and NT="long" i.e. the number of transits occurring in an assumed "long" (=150 days) pointing. Note that results from GLO scenarios around (F,G,K) stars yielded S/N<1 and are therefore not shown.

Figure 18 suggests that the lighter atmospheres with decreasing metallicity (scenarios 1-3) have increased S/N. This arose because they featured an increased atmospheric scale height (equation 3) hence a more extended atmosphere. Scenarios 4-5 which both assume x1.0 solar metallicity but with added tholin hazes have even stronger $\Delta TD$s and S/N signals because these smaller-sized hazes enhance the Rayleigh extinction feature. The pure $H_2O$ atmosphere (scenario 6) has only weak $\Delta TD$s (see 6.3). This is because firstly, steam atmospheres are less extended than lighter, hydrogen-dominated type atmospheres due to increased molecular mass and secondly, there feature strong water absorption bands in the red filter interval which reduce the value of $\Delta TD_{blue-red}$.

Figure 18 suggests for the most favorable cases (i.e. short distances of 10pc and NT="long" (number of transits occurring in 150 days) results show an increasing S/N from ~(1 to 3) for the cloud free cases with decreasing metallicity and even higher for the cases with haze. The $H_2O$ atmosphere case however lies below the detection limit. Moving to the 25pc case, Figure 18 suggests moderate S/N in the range ~ 1<S/N<2 for scenarios (3-5) and non-detection for the other scenarios. Since results for the primary $\Delta$TDs in Figure 18 imply that this detection will be challenging we do not consider the (even more challenging) occultation case.

## 7.4 Target Stars

Table 10 shows the number of stars (all sky) by distance from (1) Theoretical (black values) output from the Besançon model (Robin et al., 2003) and (2) Observations (purple values) are taken from the RECONS Astrometry Project (www.recons.org). Note that PLATO could sample up to ~half the stars in the sky (Rauer et al., 2014) (see 5.1 for mission concept):

Table 10: Estimated number of stars (all sky) by distance taken from (1) observations (shown in purple) from the RECONS Astrometry project. *RECONS does not specify numbers for M-dwarf stars over the range M0V-M5V; (2) the Hipparcos catalogue (shown in green); [#]M-dwarf stars challenging to observe with Hipparcos. §Magnitude range (2.877-4.273), (4.273-5.283), (5.283-7.45) for F, G, K stars respectively at 10pc; §§ Magnitude range(4.866-6.262), (6.262-7.276), (7.276-9.44) for F, G, K stars respectively at 25pc and (3) the Besançon model (shown in black). Values shown in red indicate stars which are difficult to observe (especially for late M-dwarfs).

| Stellar Class | RECONS <10pc | Hipparcos <10pc§ | Hipparcus <25pc§§ | Besançon <10pc | Besançon <25pc | Besançon <100pc |
|---|---|---|---|---|---|---|
| M0V-M9V | 248 | n/a[#] | n/a[#] | 299 | 5,337 | 211,337 |
| M0V-M5V | See legend* | n/a[#] | n/a[#] | 127 | 2,680 | 172,225 |
| K0V-K9V | 44 | 1-21 | 102-875 | 27 | 551 | 37,788 |
| G0V-G9V | 20 | 7-30 | 100-372 | 15 | 259 | 16,591 |
| F0V-F9V | 60 | 9-114 | 49-382 | 5 | 86 | 6,298 |

The Besançon Galaxy model (black values in Table 10) described in Robin et al. (2003) combines theoretical evolution of stars based on the initial mass function (IMF) with observational constraints from the Hipparcos mission and large scale surveys in the visual and near IR. Comparing this model with observations from Hipparcos (not shown) yielded reasonable agreement for the (F,G,K) stars but suggested an under-estimation by Hipparcos especially for the late, cool M-dwarf stars which are challenging to observe. The Gaia catalogue offers much more extensive data - but this was deemed to be not necessary for our purpose here which was to obtain an approximate estimate of the stellar fields.

Note that the RECONS observational project (purple values in Table 10) aims to map the stellar population in the solar neighborhood up to 10pc and is currently extending its survey to 25pc. Table 11 shows the total number of planets and the total transiting HJs and GLOs and non-transiting UHJs observable by PLATO on the sky:

Table 11: Total number (all sky) planets, HJs and GLOs by distance

| Type of Object | <10pc | <25pc | <100pc |
|---|---|---|---|
| Total observable planets | 174 | 3,576 | 60,677 |
| Transiting (U)HJs | 0-1 | 4 | 61 |
| UHJ phase curve | - | 28* | - |
| Transiting GLOs | 0-1 | 3** | 898** |

Planetary yields in Table 11 assume that 1% of (F,G,K) stars have (U)HJs with a 10% transit probability in the solar neighborhood (Gaudi et al., 2005) and that 33% of stars have a GLO with (3-30) $M_{earth}$ (see e.g. Narang et al., 2018 and references therein) with a 1% (**) transit probability. The Fressin et al. (2013) (Kepler) study also reported a similar GLO yield of 39%. *Assuming 1% UHJ yield and that the PLATO FCs will observe phase curves of K-stars only based on (25pc) K-star occurrence [=27 / (299+27+15+5)] in Table 10. FC coverage is 1.4% of the sky.

## 8. Conclusions

Table 12 (a-c) summarizes the number of transits ($N_{transits}$) required by PLATO to constrain (a) Hot Jupiters, (b) Ultra Hot Jupiters, and (c) the GJ 1214b-like analogues:

Table 12(a) Number of transits required by PLATO to achieve three-sigma significance (with two-sigma values shown in brackets) in the (S/N). [*]Low (high) TD = 50ppm (500ppm) in blue minus red wavelength intervals for G-star case (see 6.0). [**]Emission flux assumed to be negligible compared with reflected flux (see 6.0).

| Scenario | Distance (pc) | $N_{transits}$ K-star | $N_{transits}$ G-star | $N_{transits}$ F-star |
|---|---|---|---|---|
| HJ Primary (low TD[*]) | 25 | 47 | 28 | 17 |
| HJ Primary (low TD) | 100 | 744 (331) | 459 (205) | 278 (124) |
| HJ Primary (high TD) | 25 | 1 | 1 | 1 |
| HJ Primary (high TD) | 100 | 8 | 5 | 3 |
| HJ Reflection[**] | 25 | 3 | 6 | 31 |
| HJ Reflection[**] | 100 | 38 (17) | 94 (42) | 533 (237) |

Table 12(a) suggests that constraining HJs is achievable for all cases considered up to 25pc whereas for 100pc it becomes challenging with transmission for cases with low assumed TD and for reflection. Table 12(b) is as for Table 12 (a) but for Ultra Hot Jupiters:

Table 12(b) is as for Table 12(a) but for UHJs. [#]Assuming a blue minus red transit depth of 601ppm based on WASP-103b (see Figure 13). [##]Emission plus reflection. n/a = not applicable since it is not clear if UHJs orbiting K-stars exist in nature.

| Scenario | Distance (pc) | $N_{transits}$ K-star | $N_{transits}$ G-star | $N_{transits}$ F-star |
|---|---|---|---|---|
| UHJ Transmission[#] | 25 | 1 | 1 | 1 |
| UHJ Transmission | 100 | 8 | 5 | 3 |
| UHJ Occultation[##] | 25 | 1 | 2 | 6 |
| UHJ Occultation | 100 | 4 | 24 | 83 |
| UHJ Phase Curve | 25 | n/a | 144 (64) | 533 (237) |
| UHJ Phase Curve | 100 | n/a | 2500 (1,112) | 10,000 (4,445) |

Table 12(b) suggests that constraining UHJs is achievable for both primary transit and occultation up to 100pc but becomes challenging already at 25pc for constraining UHJ phase curves. Table 12(c) is as for Table 12(a) but for the GJ 1214b-like analogues (orbiting M4.5) with the TDs calculated from numerical atmospheric models (see 7.3):

Table 12c: as for Table 12(a) but for the GJ 1214b-like analogues (orbiting M4.5). TDs are calculated from numerical atmospheric models (7.3):

| Scenario | Distance (pc) | $N_{transits}$ |
|---|---|---|
| x1.0 Solar | 10 | 1,112 (494) |
| x0.1 Solar | 10 | 186 (83) |
| x0.01Solar | 10 | 108 (48) |
| Solar with 0.05 microns | 10 | 47 (21) |
| Solar with 0.1 microns | 10 | 29 (13) |
| x1.0 Solar | 25 | 563 (251) |
| x0.1 Solar | 25 | 1,112 (495) |
| x0.01Solar | 25 | 625 (278) |
| Solar with 0.05 microns | 25 | 278 (124) |
| Solar with 0.1 microns | 25 | 186 (83) |

In Table 12(c), steam atmosphere scenarios yielded non-achievable ($N_t$ ~several 1000s) values (not shown). Table 12(c) suggests that constraining GJ 1214b analogues is challenging but maybe achievable for a few select cases with light atmospheres in the solar neighborhood.

In summary the main conclusions of our work are:

- The PLATO FCs can constrain atmospheric properties for numerous planetary scenarios although additional information from e.g. spectroscopy is also required to address potential degeneracies.

- The PLATO FCs can detect bulk atmospheric composition and geometric albedos could be detected for (Ultra) Hot Jupiters up to ~100pc (~25pc) away assuming strong (moderate) Rayleigh slopes.

- Phase curve information is challenging but could possibly be extracted for some UHJs orbiting G stars up to 25pc away.

- For low mass low density planets, basic atmospheric types (e.g. light and cloud-free versus heavy or/and cloudy) in the upper atmosphere could be distinguished for up to a handful of cases up to~10pc away.

Therefore, while PLATO is not designed for atmospheric characterization it will nevertheless provide a list of interesting targets via its fast cameras which can then be followed up with detailed spectroscopic characterization by instruments with larger apertures such as ARIEL and JWST.

## Acknowledgements

The authors gratefully acknowledge the European Space Agency and the PLATO Mission Consortium, whose outstanding efforts have made these results possible. The authors also extend warm thanks to members of the PLATO Science Working Team (SWT) for fruitful discussion. We also thank heartily the science and local organizing committees of the PLATO atmosphere workshops held in DLR-PF Berlin as well as their scientific participants whose discussions and input considerably helped to improve this manuscript. M. G. gratefully acknowledges support by the DFG through the project GO 2610/1-1. L.C. acknowledges funding by DLR grant 50OR1804.

**Appendix A1:** Transit parameters for the PLATO FC blue and red spectral interval

| Quantity | HD 209458b | GJ 1214b (x0.01 Solar) | GJ 1214b ($H_2O$) | Earth (ADL) | Earth (Sun) |
|---|---|---|---|---|---|
| $TD_{blue} = (R_{blue}/R_*)^2$ (ppm) | 14,630[*] | 13,873[#] | 13,444[##] | 554.4[**] | 84.6[§] |
| $TD_{red} = (R_{red}/R_*)^2$ (ppm) | 14,580 | 13,483 | 13,491 | 553.4 | 84.4 |
| $TD_{blue}$ - $TD_{red}$ (ppm) | 50 | 390 | -47 | 1.0 | 0.2 |
| $\sqrt{TD_{blue}} = (R_{blue}/R_*)$ | 0.12095 | 0.11778 | 0.11595 | 0.02355 | 0.00920 |
| $\sqrt{TD_{red}} = (R_{red}/R_*)$ | 0.12075 | 0.11612 | 0.11615 | 0.02352 | 0.00918 |
| $R_*$(km)[§] | 778,971 | 143,600 | 143,600 | 695,510 | 695,510 |
| $R_{blue}$ (km) | 94,220 | 16,913.7 | 16,650.2 | 6,388.4 | 6396.8 |
| $R_{red}$ (km) | 94,059 | 16,674.3 | 16,679.2 | 6,382.7 | 6388.2 |
| $R_{blue}$ - $R_{red}$ | 161.1 | 239.4 | -29.1 | 5.8 | 8.6 |
| H (km) | 549.4 | 219.8 | 24.4 | 8.0 | 8.0 |
| $T_{eq}$ (K)[§] | 1,316 | 500 | 500 | 255 | 255 |
| g (m s$^{-2}$)[§] | 9.41 | 8.94 | 8.94 | 9.81 | 9.81 |
| Number of H observed ('n' in equation 2) | 0.29 | 1.10 | 1.20 | 0.72 | 1.08 |
| Region observed (red to blue) | 0.975 to 0.976 $R_p$ | 0.918 to 0.932 $R_p$ | 0.919 to 0.917 $R_p$ | 11.7 to 17.4km | 17.2 to 25.8km |

[§]Stellar and planetary parameters taken from the open exoplanet catalogue. Bold values are directly input into equation 2 to calculate the number of scale heights ('n' in equation 2) sampled and discussed in the text.[*]Data from Deming et al., (2013) their Figure 14. [#]Values taken from our work, see Table 8 above. [**]Taken from Scheucher et al. (2018) their Figure 7. [§] Bétrémieux, J. and Kaltenegger, L. (2013). For HD 209458b we assume $TD_{blue}$-$TD_{red}$ = 50ppm which is our standard case for HD 209458b as discussed above. For GJ 1214b we take results from model A for the x0.01 solar case (TD=390ppm) and the pure steam case (TD=-47ppm). Scale heights (H) are estimated by scaling relative to Earth (H=8km, $T_{surf}$=288K; g=9.81ms$^{-2}$) from the formula H=kT/μg (equation 3 in the main text) assuming $\mu_{Earth}$=28.8, $\mu_{HD209458b}$=2g/cm$^3$. For example: $H_{HD209458b}$=8*(28.8/2)*(1,316/288)*(9.81/9.41) =549.4km.

**Appendix A2**: TD$_{blue-red}$ ±1σ (ppm) for four example observed cases from Table 3.

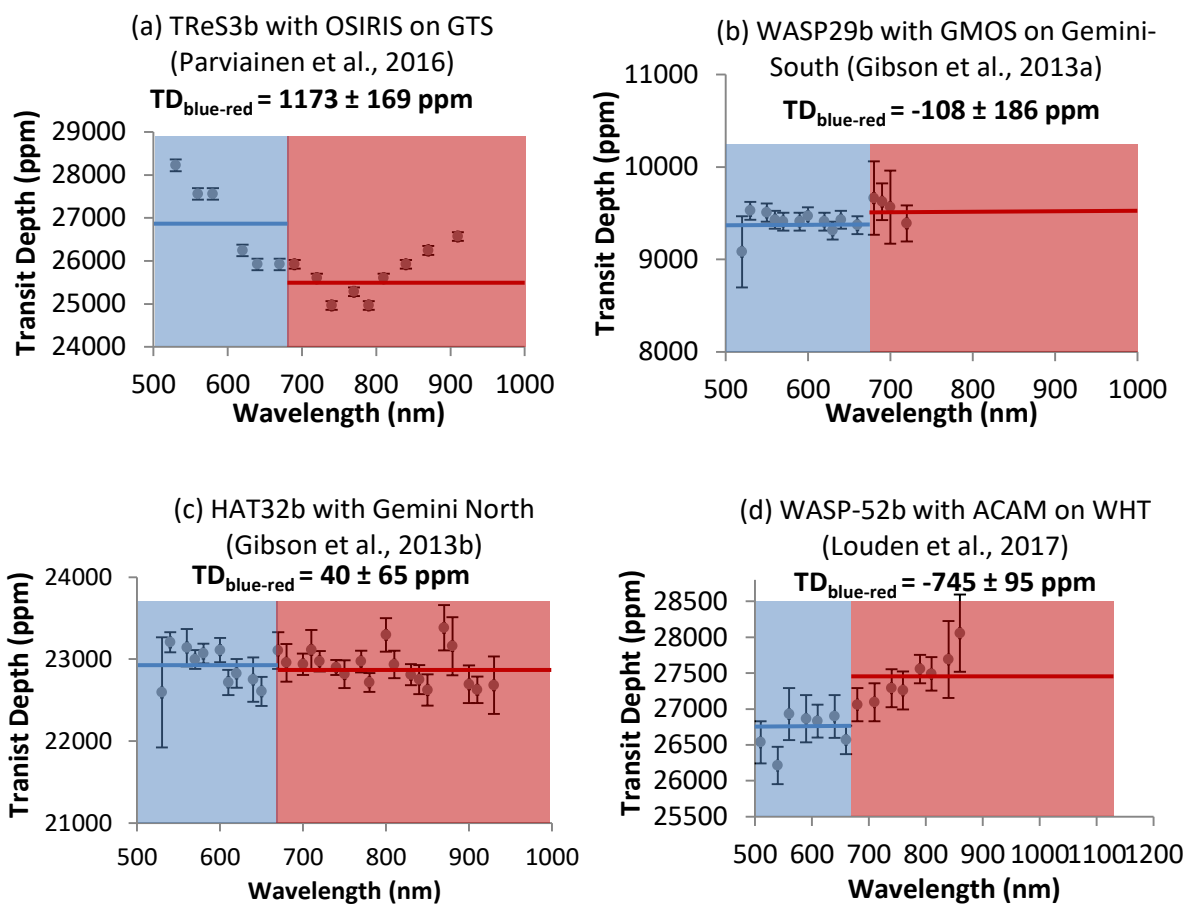

Figure A2: Evaluation of TD$_{blue-red}$ ±1σ (ppm) for four example cases from Table 3. Thick blue (red) lines show the arithmetic mean TD (ppm) values in the blue (red) interval.

Figure A2 (a-d) shows 4 cases, namely case (a) where TD$_{blue-red}$ is large, positive and statistically significant; cases b) and (c) where TD$_{blue-red}$ is small (either weakly negative as in case (b) or weakly positive as in case (c)) and non-significant, and case (d) where TD$_{blue-red}$ is large, negative and statistically significant.

**Appendix A3:** Variation of Geometric Albedo for HJ 189733b as a function of wavelength.

Data taken from polarimetry observations by Berdyugina et al. (2011).

| Wavelength Range (nm) | Geometric Albedo |
|---|---|
| (475-525) | 0.48 |
| (525-575) | 0.34 |
| (575-625) | 0.24 |
| (625-675) | 0.19 |
| (675-725) | 0.15 |
| (725-775) | 0.12 |
| (775-825) | 0.10 |
| (825-875) | 0.06 |
| (875-925) | 0.06 |
| (925-975) | 0.03 |
| (975-1025) | 0.01 |
| (1025-1075) | 0.00 |
| (1075-1125) | 0.00 |